\documentclass[amsmath,amssymb,showkeys,superscriptaddress,floatfix,aps,prd,10pt,twocolumn]{revtex4}
\usepackage{graphicx,epstopdf}
\usepackage[caption=false]{subfig}
\usepackage{multirow}
\usepackage{appendix}
\usepackage{xcolor}
\usepackage[colorlinks=true, urlcolor=blue, linkcolor=blue, citecolor=green]{hyperref}
\def\pslash{p\!\!\!\slash }
\def\qslash{q\!\!\!\slash }
\def\xslash{x\!\!\!\slash }

\begin{document}

\title{Isovector and isoscalar tensor form factors of $N(1535) \rightarrow N$ transition in light-cone QCD}
\author{Ula\c{s} \"{O}zdem}%
\email[]{ulasozdem@aydin.edu.tr}
\affiliation{Health Services Vocational School of Higher Education, Istanbul Aydin University, Sefakoy-Kucukcekmece, 34295 Istanbul, Turkey}

\begin{abstract}
We have applied isovector and isoscalar tensor current to evaluate the  tensor form factors of the $ N(1535) \rightarrow N $ transition with the help of the light-cone QCD sum rule method.  
In numerical computations, have used the most general forms of the interpolating current for the nucleon and the tensor current together with two different sets of the input  parameters in the DAs of the $N(1535)$ state. 
We have obtained that the values  of  $N(1535) \rightarrow N$ transition tensor form factors very sensitive to the input parameters of  the distribution amplitudes of the $N(1535)$ state.
We have acquired that the $Q^2$ dependence of $N(1535) \rightarrow$ transition tensor form factors is well defined by a p-pole fit function. 
\end{abstract}
 \date{\today}
\maketitle

\section{Introduction} 

The essential subject of QCD is to understand the internal structure of hadrons and their features in terms of degrees of freedom of quark-gluons. 
Hadron charges described as matrix elements of tensor, axial and vector currents between hadron states include complete knowledge about internal structure of the hadron.  At the twist-two level, the corresponding charges  are characterized by the helicity distribution $g_1(x)$, transversity distribution $h_1(x)$ and  unpolarized distribution $f_1(x)$  function of the quark.  
More generally, at the leading twist, eight generalized parton distributions (GPDs) encompass full knowledge  on the internal structure of hadrons: 
four chiral-odd spin-dependent GPDs $H_T (x, \xi, t)$, $E_T (x, \xi, t)$, $\tilde H_T (x, \xi, t)$, and $\tilde E_T (x, \xi, t)$; 
two chiral-even spin-dependent GPDs $\tilde H (x, \xi, t)$, and $\tilde E (x, \xi, t)$ and;
two chiral-even spin-independent GPDs  $H (x, \xi, t)$ and $E (x, \xi, t)$, where $\xi$ is the skewness and t = -$Q^2$ is the squared momentum transfer \cite{Ji:1996ek, Radyushkin:1997ki, Hoodbhoy:1998vm, Diehl:2001pm}. 
These observables include important knowledge about the internal structure of the hadron. They characterize, e.g., how partons are distributed in the transverse plane according to motion of the hadron or the contribution of quark orbital angular momentum to total angular momentum of the hadron. 
The helicity and unpolarized  distribution functions  can be extracted from inclusive deep-inelastic scattering data because of their chiral-even nature. 
In the forward limit, they are related to the electromagnetic, axial and pseudoscalar form factors \cite{Goeke:2001tz}.
However, the transversity distribution function, which is related to the tensor form factors in the forward limit, has chiral-odd nature, so there is a big experimental problem to measure it.
It can be acquired Drell-Yan processes and semi-inclusive deep inelastic scattering, as distributions of transversity do appear at leading twist in the cross-section.
Photo- and electro-production of mesons off the polarized nucleons and the transversely polarized Drell-Yan process are recommended as suitable ways to measure transversity distribution.
In Ref. \cite{Anselmino:2007fs}, transversity distribution of the nucleon was extracted using the experimental data from COMPASS \cite{Ageev:2006da}, HERMES \cite{Airapetian:2004tw} and Belle \cite{Abe:2005zx} Collaborations. Afterward, in Ref. \cite{Cloet:2007em} the tensor charge of the nucleon was extracted in the framework of the covariant quark-diquark model.
Moreover, tensor form factors of the nucleon have been investigated by the help of  QCD sum rule \cite{He:1994gz, He:1996wy}, axial vector meson dominance model \cite{Gamberg:2001qc}, quark model \cite{Schmidt:1997vm, Pasquini:2005dk}, chiral quark soliton model \cite{Lorce:2007fa, Ledwig:2010tu}, light-cone QCD sum rule \cite{Erkol:2011iw, Aliev:2011ku}, dihadron production \cite{Pisano:2015wnq}, lattice QCD \cite{Hagler:2007xi, Gockeler:2005cj, Bhattacharya:2015wna}, relativistic confined quark model \cite{Gutsche:2016xff}, and skyrme model \cite{Olness:1992zb}. Besides, the tensor form factors of the octet hyperons are investigated  in the framework of the chiral quark soliton model \cite{Ledwig:2010tu} and light-cone QCD sum rule \cite{kucukarslan:2016xhx}.

 Form factors play a crucial role in our comprehension of the tomography of baryons.  
The tensor form factors are missing part of the this tomography.
Recently, the measurements of exclusive electro-production of pseudoscalar mesons ($\pi$ and $\eta$ mesons) has demonstrated that these processes are responsive to chiral-odd GPDs \cite{Bedlinskiy:2012be, Bedlinskiy:2014tvi, Defurne:2016eiy, Bedlinskiy:2017yxe}.
Photo- and electro-production of pseudoscalar mesons can be used to extract the tensor form factors of baryons \cite{Bedlinskiy:2012be}.  
In the short run, remarkably more accurate measurements of the nucleon tensor form factors are expected at Jefferson Laboratory (JLab) by the CLAS Collaboration.
Besides, the experiments designed at CLAS Collaboration, have been aimed to investigate features of electro-excitaton of nucleon resonances in photo- and electro-production reactions \cite{Aznauryan:2012ba}.
Inspired by the future experiments at JLab,
%
we aim to investigate the isovector and isoscalar tensor form factor $N(1535) \rightarrow N$ transition up to a momentum transfer of $Q^2  \leq$ 10 GeV$^2$  with the help of the light-cone QCD sum rule. 
To our knowledge, this is the first study in the literature
committed to the examination of the $N(1535) \rightarrow N$ transition tensor form factors.
 In the light-cone QCD sum rule method, the hadronic observables are described in connection with the properties of the vacuum and distribution amplitudes (DAs) of the hadrons under the investigation \cite{Braun:1988qv, Balitsky:1989ry, Chernyak:1990ag}. Since the hadronic observables are described in connection with the features of the QCD vacuum and the DAs, any ambiguity in these variables reflects to the ambiguity of the predictions of the hadronic observables.
 Note that the electromagnetic \cite{Anikin:2015ita}, axial \cite{Aliev:2019tmk} and gravitational \cite{ozdem:2019pkg} form factors for $N(1535) \rightarrow N$ transition have been evaluated with the help of light-cone QCD sum rule.

 This article is organized in the following manner: In Sec. \ref{secII} we present the details of our light-cone QCD sum rule calculations.
 In Sec. \ref{secIII} we analyze the obtained results and give our conclusions.

 \section{Isovector and isoscalar tensor Form Factors of  $N(1535) \rightarrow N$ transition }\label{secII}

 The matrix element of the isovector and isoscalar tensor current between nucleon and $N(1535)$ baryons is defined by three dimensionless invariant form factors as presented  \cite{Hagler:2009ni, Gockeler:2006zu}
 \begin{widetext}
  \begin{align}\label{matFFs}
\langle N(p^\prime)|J_{\mu\nu}|N(1535)(p)\rangle =
 \bar{u}(p^\prime)\Bigg[i\sigma_{\mu\nu} H_T^{I=0,1}(Q^2) 
  +\frac{\gamma_\mu q_\nu-\gamma_\nu q_\mu}{2 \bar m} E_T^{I=0,1}(Q^2)
  +\frac{\tilde P_\mu q_\nu-\tilde P_\nu q_\mu}{2 \bar m^2} \tilde{H}_T^{I=0,1}(Q^2) \Bigg] \gamma_5 u(p),
\end{align}
\end{widetext}
 where $\bar m$ =$ ( m_N+m_{N(1535)})/2$,  $\sigma_{\mu\nu}= \frac{i}{2}[\gamma_\mu, \gamma_\nu]$,  $q = p-p'$, 
$\tilde P= p'+p$ and, $F^{I=1} = F^u-F^d$ and  $F^{I=0} = F^u+F^d$  for any of the form factors, $F $= $E_T$, $H_T$  or $\tilde{H}_T$.

To derive the light-cone QCD sum rules for isovector and isoscalar tensor form factors of $N(1535) \rightarrow N$ transition, 
we consider subsequent correlator for our analysis
\begin{align}\label{corf}
\Pi_{\mu\nu}(p,q)=i\int d^4 x e^{iqx} \langle 0 |\mathcal{T}[J_{N}(0)J_{\mu \nu}(x)]|N(1535)(p)\rangle,
\end{align}
where $J_{\mu\nu}(x)$ is the tensor current and 
$J_N(0) $ are interpolating currents for nucleon states. 
The explicit forms of the $J_N(0) $ and $J_{\mu\nu}(x)$ are given as
\begin{align}
J_N(0) &=2\epsilon^{abc}\sum_{\ell=1}^{2}(u^{aT}(x) C J_1^\ell u^b(x))J_2^\ell d^c(x),\nonumber\\
J_{\mu\nu}(x)&=\bar{u}^d(x)i\sigma_{\mu\nu} u^d(x) \pm \bar{d}^e(x)i\sigma_{\mu\nu} d^e(x),
\end{align}
respectively, where $J_1^1=I$, $J_1^2=J_2^1=\gamma_5$,  $J_2^2=t$, which is an arbitrary parameter that fixes the mixing of two local operators, and C denotes charge conjugation.

To acquire the sum rules for isovector and isoscalar tensor form factors of the $N(1535) \rightarrow N$ transition 
the correlator in Eq.(\ref{corf}) is obtained from the subsequent three steps:\\
$\bullet$ The correlator is saturated by complete set of hadronic states,  which are have the same quantum numbers as interpolating currents (hadronic representation),\\
$\bullet$ The correlator is obtained in connection with quark and gluon degrees of freedom interacting with non-perturbative QCD vacuum (QCD representation).\\
$\bullet$ Then match these two independent representations of the correlator to one another employing the quark-hadron
duality ansatz. To keep under control undesirable contributions coming from the higher and excited states, we perform a Borel
transformation, in addition to continuum subtraction to both representations of the obtained corresponding sum rules.

As we mentioned above in order to evaluate the correlator in connection with hadron features, a complete hadronic set with the same quantum numbers as the interpolation currents is inserted. After that, the correlation function becomes
 \begin{align}\label{phys}
 \Pi_{\mu\nu}^{Had}(p,q) =&\sum_{s{'}} \frac{\langle0|J_N(0)|{N(p',s')}\rangle 
 }{m^2_{N}-p'^2}\nonumber\\
&
 \langle {N(p',s')} |J_{\mu \nu}(x)|N(1535)(p,s)\rangle +...,
\end{align}
where 
\begin{align}
\langle0|J_N(0)|{N(p',s')}\rangle &= \lambda_N u_{N}(p',s'),\label{res}
\end{align}
with $ \lambda_N $ and $ u_{N}(p',s') $ being the residue and Dirac spinor of nucleon, respectively. Summation over the spins of the nucleon is performed as
\begin{align}
\sum_{s'} u_N(p',s')\bar u_N(p',s') &= \pslash' + m_N.\label{spinor}
\end{align}

Substituting Eqs. (\ref{matFFs}),  (\ref{res}) and (\ref{spinor}) into Eq. (\ref{phys}), we acquire the
 correlator in the way of the hadronic features as

\begin{widetext}
\begin{align}\label{hads}
  \Pi_{\mu\nu}^{Had}(p,q) = \frac{ \lambda_N}{m_N^2-p'^{2}}(\pslash^{\prime}+m_N)
\left[i\sigma_{\mu\nu} H_T^{I=0,1}(Q^2) +\frac{\gamma_\mu q_\nu-\gamma_\nu q_\mu}{2 \bar m} E_T^{I=0,1}(Q^2)
 +\frac{\tilde P_\mu q_\nu-\tilde P_\nu q_\mu}{2 \bar m^2} \tilde{H}_T^{I=0,1}(Q^2) \right] \gamma_5\, u(p) .
 \end{align}
\end{widetext}

The next step is to evaluate the correlator in Eq. (\ref{corf}) with respect to quarks and gluon properties in
deep Euclidean region. Employing the expression for $J_N(0)$ and $J_{\mu\nu}(x)$ and Wick's theorem, 
the QCD representation of the correlator is obtained as
\begin{widetext}
\begin{align}\label{corrfunc}
	\Pi_{\mu\nu}^{QCD}(p,q)&=-\int d^4 x e^{iqx}\Bigg[\bigg\{ (\gamma_5)_{\gamma\delta}\, C_{\alpha\beta}\, (i\sigma_{\mu\nu})_{\omega \rho} 
	+
	t\, (I)_{\gamma\delta}\,(C \gamma_5)_{\alpha\beta}\,( i \sigma_{\mu\nu})_{\omega \rho}	\bigg\}\nonumber\\
   & \times	\bigg\{ \langle 0|\epsilon^{abc} u_{\sigma}^a(0) u_{\theta}^b(x) d_{\phi}^c(0)|N(1535)(p)\rangle
      \Big (\delta_\sigma^\alpha \delta_\theta^\rho \delta_\phi^\beta S_u(-x)_{\delta \omega}
     +\, \delta_\sigma^\delta \delta_\theta^\rho \delta_\phi^\beta S_u(-x)_{\alpha \omega}\Big)
     \nonumber\\
     & \pm \,\langle 0|\epsilon^{abc} u_{\sigma}^a(0) u_{\theta}^b(0) d_{\phi}^c(x)|N(1535)(p)\rangle
\, \delta_\sigma^\alpha \delta_\theta^\delta \delta_\phi^\rho S_d(-x)_{\beta \omega} 
    \bigg\}
   \Bigg],
\end{align}
\end{widetext}
where $S_q(x)$ is the light-quark propagator  and it is given as ($m_q =0$)
\begin{align}
\label{edmn09}
S_{q}(x) &= 
i\frac{\xslash}{2 \pi^2 x^4}
- \frac{\langle  \bar qq \rangle }{12}
- \frac{\langle \bar q \sigma.G q \rangle }{192}x^2 
\nonumber\\
&
-\frac {i g_s }{32 \pi^2 x^2} ~G^{\mu \nu} (x) \bigg[\rlap/{x}
\sigma_{\mu \nu} +  \sigma_{\mu \nu} \rlap/{x}
 \bigg].
\end{align}
The $\langle 0|\epsilon^{abc} u_{\sigma}^a(x_1) u_{\theta}^b(x_2) d_{\phi}^c(x_3)|N(1535)(p)\rangle $
matrix element in Eq. (\ref{corrfunc}) is can be written in terms of the DAs of the $N(1535)$ state  and  it is necessary for further computations. 
The comprehensive expression of this matrix elements are presented in Ref.~\cite{maxiphd}. After employing the explicit forms of the above matrix elements and the light-quark propagator, we acquire expressions in x-space. 
Then we apply Fourier transforms to transfer these expressions into the momentum space.

The desired light-cone sum rules are obtained by matching both representations of the correlation function. In order to do this, we have to choose different and independent Lorentz structures. For this purpose, we choose  $p_\mu q_\nu \gamma_5$, $p_\mu \gamma_\nu \gamma_5$ and $p_\mu q_\nu \qslash \gamma_5$ structures for $E_T^{I=0,1}(Q^2)$, $H_T^{I=0,1}(Q^2)$  and $\tilde{H}_T^{I=0,1}(Q^2)$ form factors, respectively. As a result, we get the light-cone sum rules 
\begin{widetext}
\begin{align}
E_T^{I=1}(Q^2)  \frac{ \lambda_N}{m_N^2-p'^{2}} &= \bar{m}\, \Pi_1^{QCD}, ~~~~~~~~~~~~~
E_T^{I=0}(Q^2)  \frac{ \lambda_N}{m_N^2-p'^{2}} = \bar{m}\, \Pi_2^{QCD},\label{ETFF1}\\
H_T^{I=1}(Q^2)  \frac{ \lambda_N}{m_N^2-p'^{2}} &= -\frac{1}{2}\, \Pi_3^{QCD}, ~~~~~~~~~~~
H_T^{I=0}(Q^2)  \frac{ \lambda_N}{m_N^2-p'^{2}} = -\frac{1}{2}\, \Pi_4^{QCD},\label{HTFF1}\\
\tilde{H}_T^{I=1}(Q^2)  \frac{ \lambda_N}{m_N^2-p'^{2}} &= -\bar{m}^2\, \Pi_5^{QCD}, ~~~~~~~~~
\tilde{H}_T^{I=0}(Q^2)  \frac{ \lambda_N}{m_N^2-p'^{2}} = -\bar{m}^2\, \Pi_6^{QCD}. \label{HtilTFF1},
\end{align}
\end{widetext}

The $\Pi_i^{QCD}$ functions appearing in Eqs. (\ref{ETFF1})-(\ref{HtilTFF1}) are quite long and not illuminating. However, as an example, we give the result
of the $\Pi_1^{QCD}$. The remaining five of these functions have more or less similar forms.
\begin{widetext}
\begin{align}
\Pi_1^{QCD}&= 2\,m^2_{N(1535)}\Bigg\{\int_0^1 d\alpha \frac{\alpha}{(q-p\alpha)^4} \int_{\alpha}^1dx_2 \int_0^{1-x_2}dx_1\Bigg[ (1-t)[2A_1-2A_2-A_3+3A_4+2V_1+2V_2-4V_3\nonumber\\
&+2V_4-4V_5]+(1+t)[2P_1-2P_2+2S_1-2S_2+2T_2+4T_3-6T_5-4T_7]\Bigg] (x_1,x_2,1-x_1-x_2)\nonumber\\
&+2\,\int_0^1 d\alpha \frac{\alpha}{(q-p\alpha)^4} \int_{\alpha}^1dx_3 \int_0^{1-x_3}dx_1\Bigg[(1+t)[-P_1+P_2-S_1+S_2-T_1+T_5+T_7
+T_8]\Bigg]\nonumber\\
&\times (x_1,1-x_1-x_3,x_3)\nonumber\\
&+2\,\int_0^1 d\beta \int_{\beta}^1 d\alpha \frac{1}{(q-p\beta)^4} \int_{\alpha}^1dx_2 \int_0^{1-x_2}dx_1\Bigg[(1+t)[-T_2+T_3+T_4-T_5-T_7-T_8]\Bigg] \nonumber\\
&\times (x_1,x_2,1-x_1-x_2)\nonumber\\
&-2\,\int_0^1 d\beta \int_{\beta}^1 d\alpha \frac{1}{(q-p\beta)^4} \int_{\alpha}^1dx_3 \int_0^{1-x_3}dx_1\Bigg[(1+t)[T_2-T_3-T_4+T_5+T_7+T_8]\Bigg] \nonumber\\
&\times (x_1,1-x_1-x_3,x_3)\nonumber\\
&+8\,m^2_{N(1535)}\int_0^1 d\beta \int_{\beta}^1 d\alpha \frac{\beta^2}{(q-p\beta)^6} \int_{\alpha}^1dx_2 \int_0^{1-x_2}dx_1\Bigg[(1+t)[-T_2+T_3+T_4-T_5-T_7-T_8]\Bigg] \nonumber\\
&\times (x_1,x_2,1-x_1-x_2)\nonumber\\
&-8\,m^2_{N(1535)}\int_0^1 d\beta \int_{\beta}^1 d\alpha \frac{\beta^2}{(q-p\beta)^6} \int_{\alpha}^1dx_3 \int_0^{1-x_3}dx_1\Bigg[(1+t)[T_2-T_3-T_4+T_5+T_7+T_8]\Bigg] \nonumber\\
&\times (x_1,1-x_1-x_3,x_3)\Bigg\},
\end{align}
\end{widetext}
where, $ A_i $, $ P_i $, $ V_i $, $ S_i $  and  $ T_i $ are distribution amplitudes of different twists. They have been expressed with respect to $N(1535)$ state wavefunctions. The explicit forms of these wavefunctions are presented in Ref.~\cite{maxiphd}.
To eliminate contributions coming from the excited and continuum states the Borel transformation and continuum subtraction are performed. 
The suppression of the excited and continuum states can be accomplished by means of the subsequent subtraction rules~\cite{Braun:2006hz}:
\begin{widetext}
\begin{align}
		\int dz \frac{\rho(z)}{(q-zp)^2}\rightarrow &-\int_{x_0}^1\frac{dz}{z}\rho(z) e^{-s(z)/M^2}, \nonumber		\\
		\int dz \frac{\rho(z)}{(q-zp)^4}\rightarrow & \frac{1}{M^2} \int_{x_0}^1\frac{dz}{z^2}\rho(z) e^{-s(z)/M^2}
		+\frac{\rho(x_0)}{Q^2+x_0^2 m^2_{N}} e^{-s_0/M^2},\nonumber\\
		\int dz \frac{\rho(z)}{(q-zp)^6}\rightarrow & -\frac{1}{2M^4}\int_{x_0}^1\frac{dz}{z^3}\rho(z) e^{-s(z)/M^2}
		-\frac{1}{2M^2}\frac{\rho(x_0)}{x_0(Q^2+x_0^2m_N^2)}e^{-s_0/M^2}\nonumber\\
		&+\frac{1}{2}\frac{x_0^2\,e^{-s_0/M^2}}{Q^2+x_0^2m_N^2}\bigg[\frac{d}{dx_0}\frac{\rho(x_0)}{x_0(Q^2+x_0^2m_N^2)}\bigg],
	\label{subtract3}
\end{align}
\end{widetext}
where
\begin{align}
s(z) =& (1-z)m^2_{N}+\frac{1-z}{z}Q^2,\nonumber\\
x_0 =& \frac{\sqrt{(Q^2+s_0-m^2_{N})^2+4 m^2_{N} Q^2}-(Q^2+s_0-m^2_{N})}{2m^2_{N}}.
\end{align}
The residue of the nucleon, $\lambda_N$, is needed for the numerical computation of $N(1535) \rightarrow N$ transition tensor form factors. 
The $\lambda_N$ is specified from two-point QCD sum rules \cite{Aliev:2011ku}:
\begin{align}
\label{residue}
\lambda_N &=\Bigg[e^{\frac{m_N^2}{M^2}}\bigg\{\frac{M^6}{256 \pi^4} (5+2 t + t^2) E_2(y) \nonumber\\
&- \frac{\langle \bar{q}q \rangle^2}{6}  \bigg( 6 (1-t^2)-(1-t)^2  \nonumber\\
&- \frac{m_0^2 }{4 M^2} \Big[12 (1-t^2) 
- (1-t)^2  \Big] \bigg)\Bigg\}\Bigg]^{1/2}, 
\end{align}
where
\begin{eqnarray}
\label{nolabel}
y &=& s_0/M^2,\nonumber 
\end{eqnarray}
and 
\begin{eqnarray}
E_n(y)&=&1-e^{-y}\sum_{i=0}^{n}\frac{y^i}{i!}~. \nonumber
\end{eqnarray}

\section{Numerical analysis and conclusion}\label{secIII}

In this section, we have obtained numerical analysis of $ N(1535) \rightarrow N $ transition tensor form factors.
%
 The DAs of $N(1535)$ state have been evaluated by means of the light-cone QCD sum rule in Ref. \cite{maxiphd}. 
 The numerical values of the input parameters inside the DAs of the $N(1535)$ state are given in  Table \ref{tab:data}, which are obtained at  renormalization scale  $\mu^2 = 	2.0~\mathrm{GeV}^2$. Furthermore, we use  $ \lambda_1^N m_N = -3.88 (2)(19)\times 10^{-2}$ GeV$^3$ and  $\lambda_2^{N(1535)}  m_{N(1535)} = 8.97 (45)\times 10^{-2}$ GeV$^3$, given in Ref. \cite{Braun:2014wpa} at renormalization $\mu^2 = 4.0$ GeV$^2$, by rescaling to  $\mu^2 =2.0~\mathrm{GeV}^2$.
 Beside these values, we use  $m_{N(1535)} = 1.51 \pm 0.01$ GeV, $m_N = 0.94$ GeV~\cite{Tanabashi:2018oca},  $m_0^2=0.8 \pm 0.1$~GeV$^2$ and 
  $\langle \bar{q}q\rangle=(-0.24\pm 0.01)^3$~GeV$^3$~\cite{Ioffe:2005ym}.

 \begin{widetext}

   \begin{table}[htp]
\caption{
 Input parameters of the $N(1535)$ state DAs for the two different sets.
}\label{tab:data}
\renewcommand{\arraystretch}{1.3}
\addtolength{\arraycolsep}{-0.5pt}
\small
$$
\begin{array}{|c|c|c|c|c|c|c|c|c|c|}
\hline \hline
\mbox{Model} &\mid \lambda_1^{N(1535)}/\lambda_1^N \mid &
f_{N(1535)}/\lambda_1^{N(1535)} & \varphi_{10} & \varphi_{11} & \varphi_{20} &
\varphi_{21} & \varphi_{22} & \eta_{10} & \eta_{11} \\  \hline
\mbox{ LCSR-I} & 0.633 & 0.027 & 0.36 & -0.95 & 0 & 0 & 0 & 0     & 0.94 \\
\mbox{LCSR-II}  & 0.633 & 0.027 & 0.37 & -0.96 & 0 & 0 & 0 & -0.29 & 0.23 \\
\hline \hline
\end{array}
$$
\renewcommand{\arraystretch}{1}
\addtolength{\arraycolsep}{-1.0pt}
\end{table}

\end{widetext}

The predictions for the isovector and isoscalar tensor form factors depend on three auxiliary parameters; the Borel mass
parameter $M^2$, arbitrary mixing parameter $t$ and continuum threshold $s_0$. 
For the quality of the numerical values of the physical observables, we should minimize the dependence of the results on these parameters.
The $M^2$ can change in the interval that the results relatively weakly depend on it with respect to the standard definition. 
The upper limit of it is acquired by demanding the maximum pole contributions and 
its lower limit is acquired from the convergence of the operator product expansion and exceeding of the perturbative part over nonperturbative contributions. 
 The $t$ is chosen such that, the estimations of the isovector and isoscalar tensor form factors are reasonably insensitive of the
values of $t$.
The working region for the $s_0$ is chosen such that the maximum pole contribution is obtained and the results relatively weakly depend on its choices. 
These constraints lead to the working intervals for auxiliary parameters as
  \begin{align}
   & 2.50~\mathrm{GeV}^2 \leq M^2 \leq  3.50~\mathrm{GeV}^2,\nonumber\\
   & 2.50~\mathrm{GeV}^2\leq s_0 \leq  3.00~\mathrm{GeV}^2, \nonumber \\
   &~~~ -3.00 ~ \leq  ~t~ \leq ~ -5.00.\nonumber
  \end{align}
  
  In Figs. \ref{isovectorMsqfigs} and \ref{isoscalarMsqfigs}, we show dependency of isovector and isoscalar tensor form factors with respect to the Borel mass parameter at three fixed values of the continuum threshold and two fixed values of the arbitrary mixing parameter in their working interval.
 The results show good stability against the variations of the Borel mass parameters, as desired. 
In Figs. \ref{isovectorQsqfigs} and \ref{isoscalarQsqfigs}, we plot the dependence of the isovector and isoscalar tensor form factors
on $Q^2$ for various values of $s_0$ and $t$ in their working regions and at the fixed values of $M^2$ = 3.00 GeV$^2$ for LCSR-I and LCSR-II 
values of input parameters entering the DAs.
All the form factors taken into account show a similar dependence on $Q^2$ for LCSR-I and LCSR-II except the form factor 
$\tilde H_T^{I=0}$($Q^2$). This form factor changes its sign in the region under consideration, so its results are not given in the text.
We should note here that  the light-cone QCD sum rule approach is trustworthy only for $Q^2 > 1.0$ GeV$^2$. 
On the other hand, the baryon mass corrections of the DAs $\sim m^2/Q^2$ become very large for $Q^2 < 2.0$ GeV$^2$, in other words the light-cone QCD sum rules turn out to be untrustworthy.
Thus, we expect the light-cone QCD sum rule to be more reliable and effective in the region of  2.0 GeV $^2 \leq Q^2 \leq $ 10.0 GeV$^2$.

As we mentioned above our sum rules work only for $Q^2 \geq $ 2.0 GeV$^2$. However, we want to extend our analysis to the region  $0 \leq Q^2<2$. 
To do this, some fit parameters need to be used.
Our numerical investigations indicate that the isovector and isoscalar tensor form factors of $N(1535) \rightarrow N$ transition are nicely defined by employing the p-pole fit function:
 \begin{align}
{\cal F}(Q^2)= \frac{{\cal F}(0)}{\Big(1+ Q^2/(p\,m^2_{p})\Big)^p}.
\end{align} 

The numerical results obtained for $N(1535) \rightarrow N$ transition isovector and isoscalar tensor form factors are given in Table \ref{fit_table}.
The results obtained by employing LCSR-I and LCSR-II parameters were found to be quite different from each other.
The numerical values of the form factors  $E_T^{I=0,1}$($Q^2=0$), $H_T^{I=0,1}$($Q^2=0$) and $\tilde H_T^{I=0,1}$($Q^2=0$)
for the LCSR-II numerical values are smaller than those for the LCSR-I parameters.
 As one can see from Table \ref{tab:data}, the essential difference between input parameters of the DAs is the numerical values for the $\eta_{10}$ and $\eta_{11}$, which are related to the p-wave three-quark wave functions of the $N(1535)$ state and, therefore to the distribution of orbital angular momentum.
This means these form factors are very sensitive to the shape parameters of the DAs of the $N(1535)$ state that parametrize relative orbital angular momentum of the quarks.
\begin{widetext}

\begin{table}[htp]
\caption{ The obtained numerical values for the parameters of the isovector and isoscalar tensor form factors by employing the p-pole fit functions.
}
\hspace*{-0.5cm}
\begin{tabular}{ |l|c|c|c|c|c|c|}
\hline\hline
&\multicolumn{3}{|c|}{LCSR-I} &\multicolumn{3}{|c|}{LCSR-II}\\
\hline\hline
\begin{tabular}{c}Form factors \end{tabular}& \begin{tabular}{c} ${\cal F}(0)$ \end{tabular} & \begin{tabular}{c}$m_{p}$(GeV) \end{tabular}& \begin{tabular}{c}p \end{tabular}
&${\cal F}(0)$ & $m_{p}$(GeV) & p \\ \hline\hline
        $E_T^{I=1}$($Q^2$)      &$7.54\pm 1.26$  &$1.10 \pm 0.05 $ &$3.6-4.0$ & $3.48 \pm 0.84$ &$1.07 \pm 0.07$ &$3.8-4.2$ \\
        $E_T^{I=0}$($Q^2$)      &$5.05\pm 1.01$  &$1.13 \pm 0.08 $ &$3.6-4.0$ & $3.00 \pm 0.66$ &$1.14 \pm 0.10$ &$3.6-4.0$ \\
        $H_T^{I=1}$($Q^2$)      &$5.22\pm 0.27$  &$1.30 \pm 0.10 $ &$3.0-3.4$ & $1.51 \pm 0.20$ &$1.28 \pm 0.10$ &$3.0-3.4$ \\
        $H_T^{I=0}$($Q^2$)      &$3.37\pm 0.47$  &$1.28 \pm 0.11 $ &$3.0-3.4$ & $1.10 \pm 0.20$ &$1.32 \pm 0.10$ &$3.0-3.4$ \\
        $\bar{H}_T^{I=1}$($Q^2$) &$14.51\pm 4.43$  &$1.02 \pm 0.10 $ &$3.6-4.0$ & $1.30 \pm 0.39$ &$1.18 \pm 0.14$ &$3.6-4.0$ \\
        $\bar{H}_T^{I=0}$($Q^2$) &$-$  &$-$ &$-$ & $-$ &$-$ &$-$ \\
\hline \hline
\end{tabular}
	\label{fit_table}
\end{table}

\end{widetext}
In summary, we have applied isovector and isoscalar tensor current to evaluate the  tensor form factors of the $ N(1535) \rightarrow N $ transition with the help of the light-cone QCD sum rule method.  
In numerical computations, we have used the most general forms of the interpolating current for the nucleon and the tensor current together with two different sets of the input  parameters in the DAs of the $N(1535)$ state. 
We have obtained that the values  of  $N(1535) \rightarrow N$ transition tensor form factors are very sensitive to the input parameters of  the DAs of the $N(1535)$ state.
We have acquired that the $Q^2$ dependence of $N(1535) \rightarrow$ transition tensor form factors are well defined by a p-pole fit function. 
To our knowledge, this is the first study in the literature committed to the examination of the  $ N(1535) \rightarrow N $ transition
tensor form factors. Thus, experimental data or theoretical predictions are not yet available to compare our numerical results with them.
A comparison of the results acquired with the estimations of other theoretical approximations, such as the quark model, chiral
perturbation theory, lattice QCD, etc., would also be interesting.

\section{Acknowledgments}
The author is thankful to K. Azizi for helpful remarks, comments and discussions.

\bibliography{2-refs}

\begin{widetext}
 
 \begin{figure}[t]
\centering
  \subfloat[]{ \includegraphics[width=0.35\textwidth]{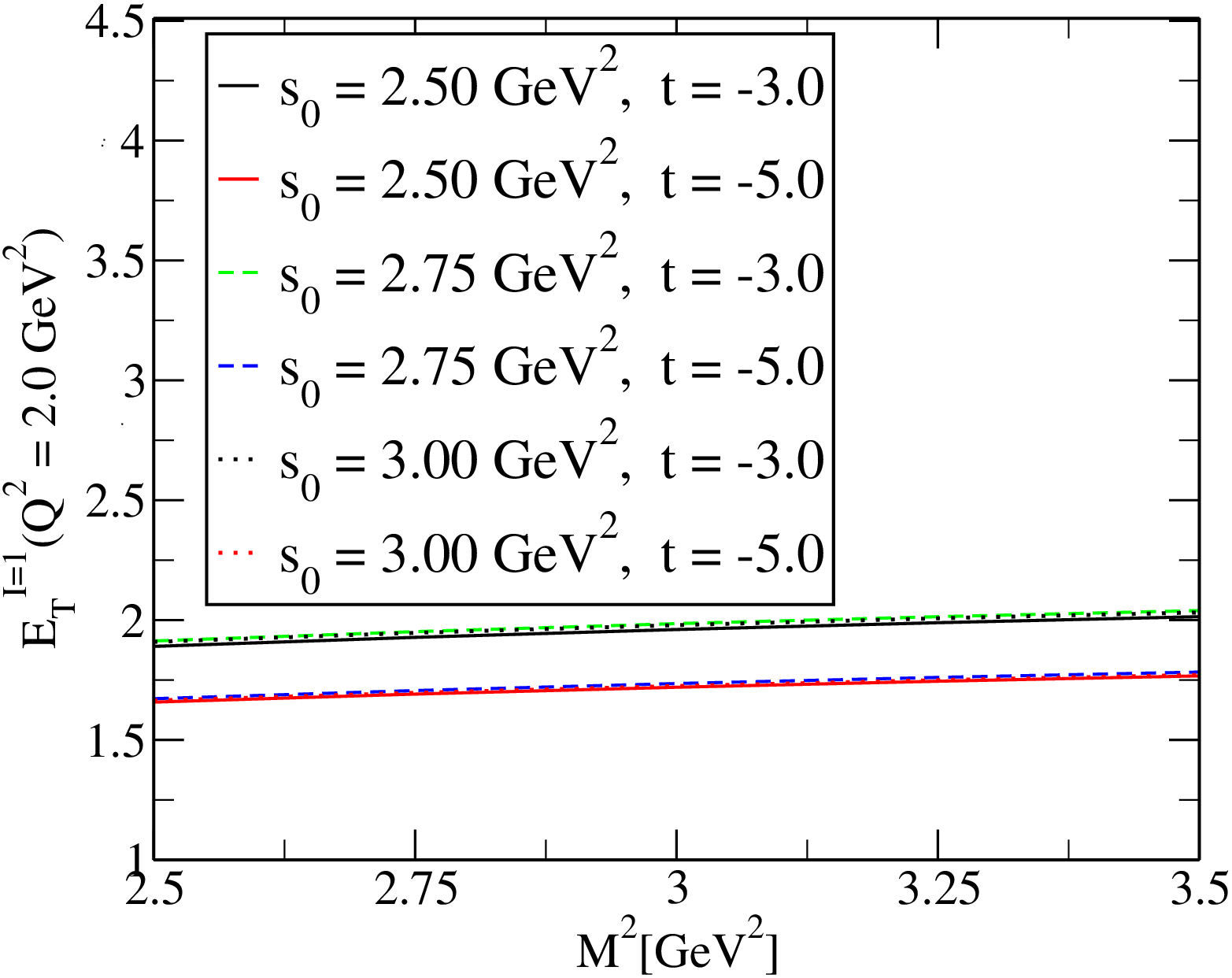}}~~~
 \subfloat[]{ \includegraphics[width=0.35\textwidth]{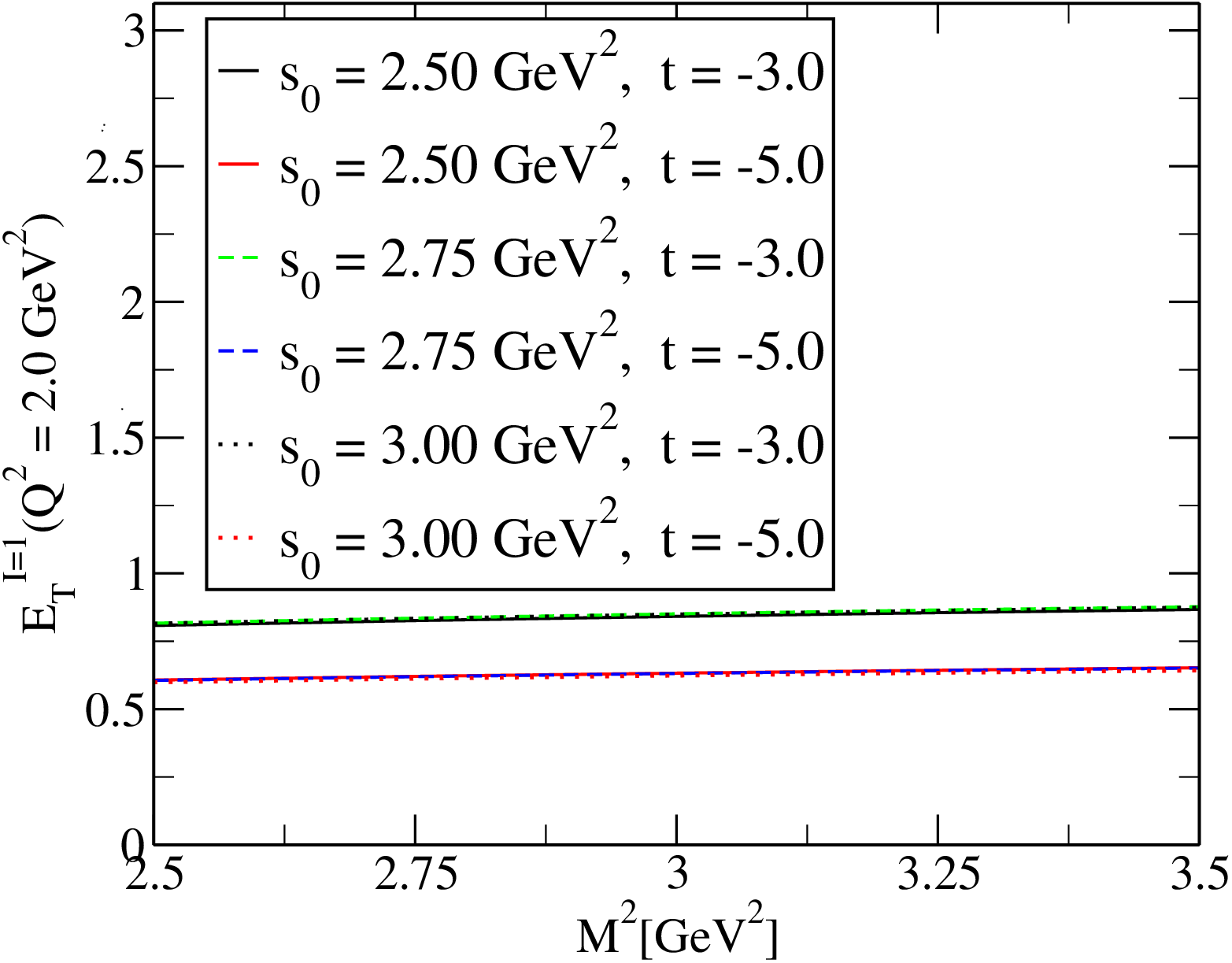}}~~~\\
  \subfloat[]{ \includegraphics[width=0.35\textwidth]{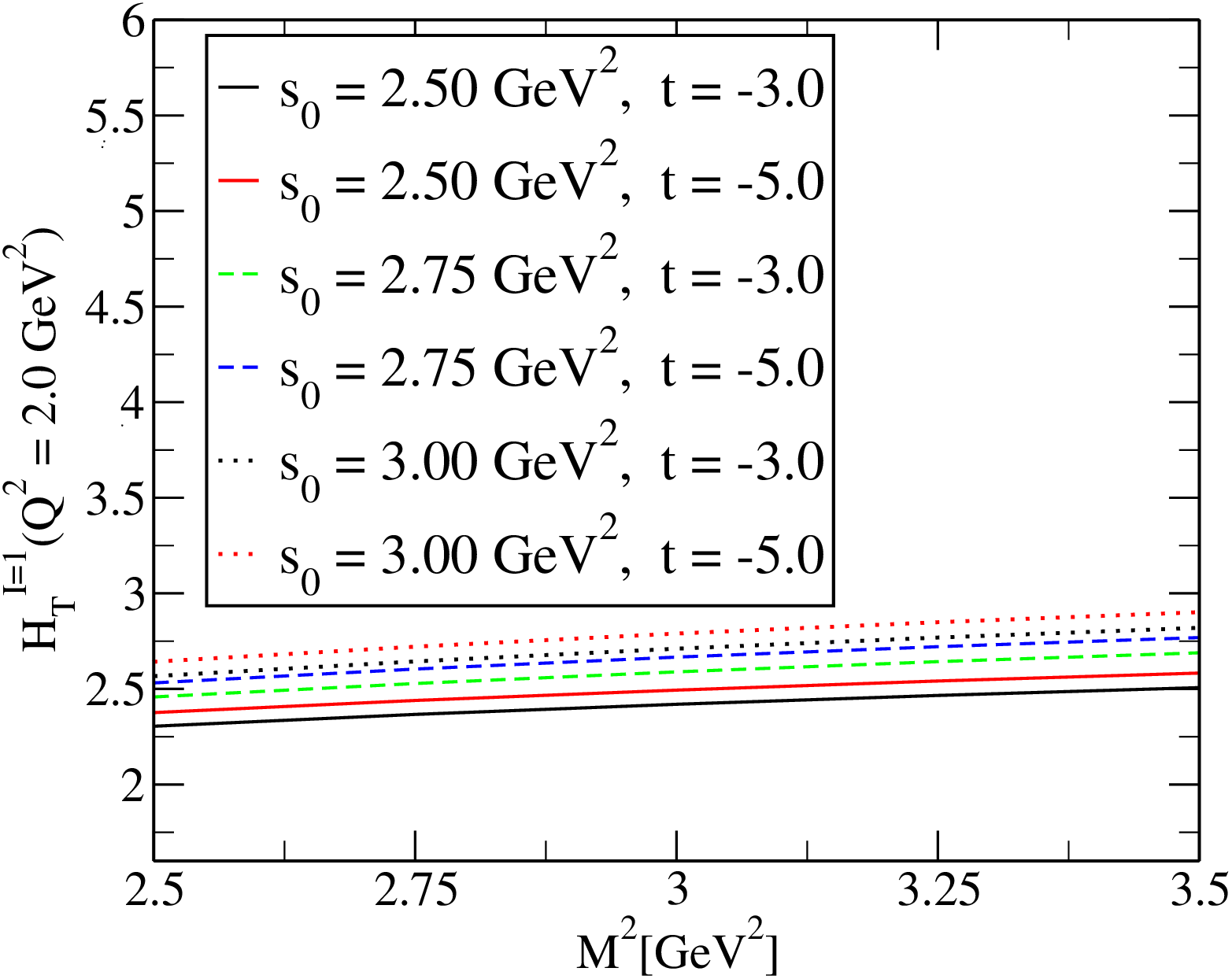}}~~~
 \subfloat[]{ \includegraphics[width=0.35\textwidth]{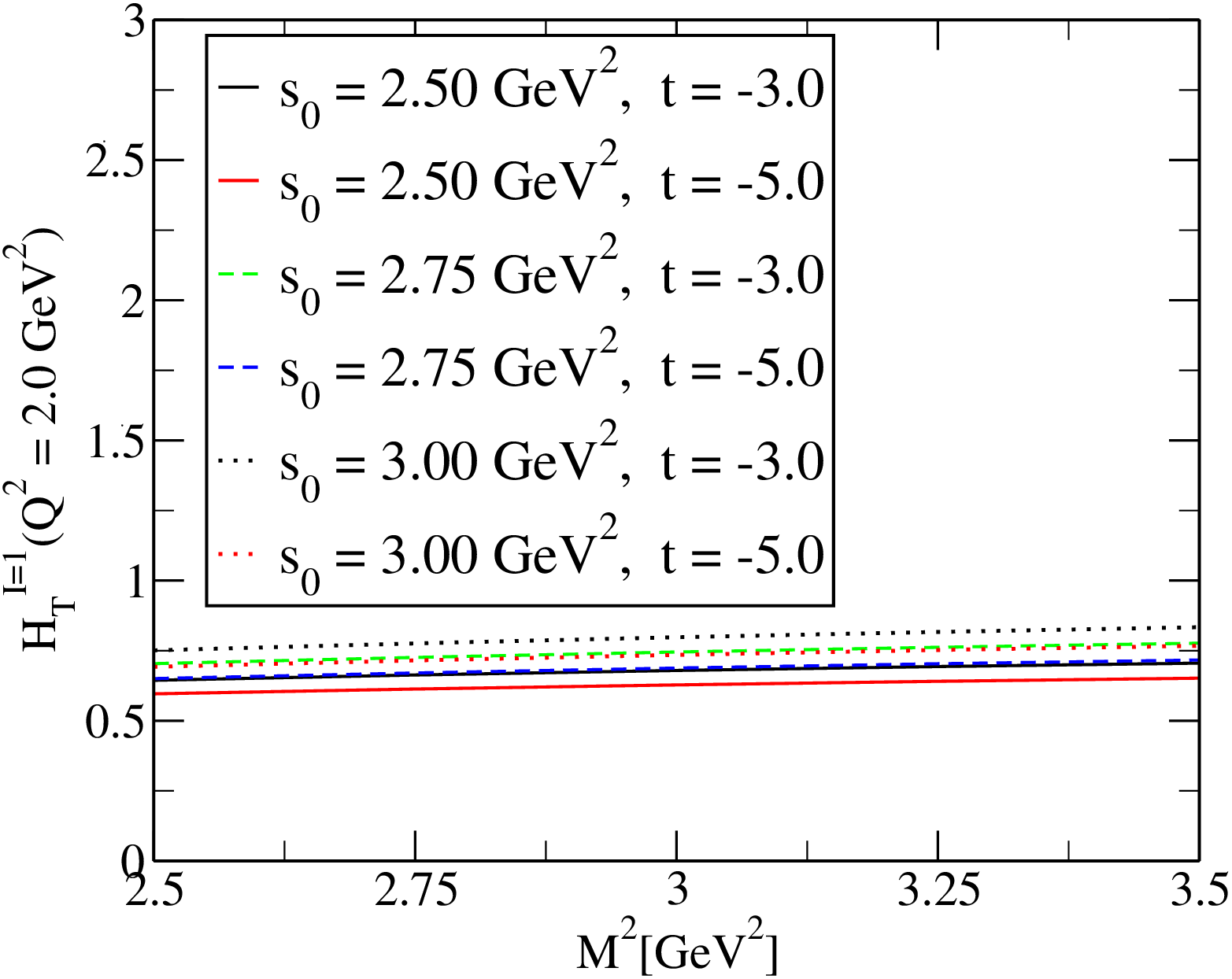}}~~~\\
  \vspace{0.2cm}
   \subfloat[]{ \includegraphics[width=0.35\textwidth]{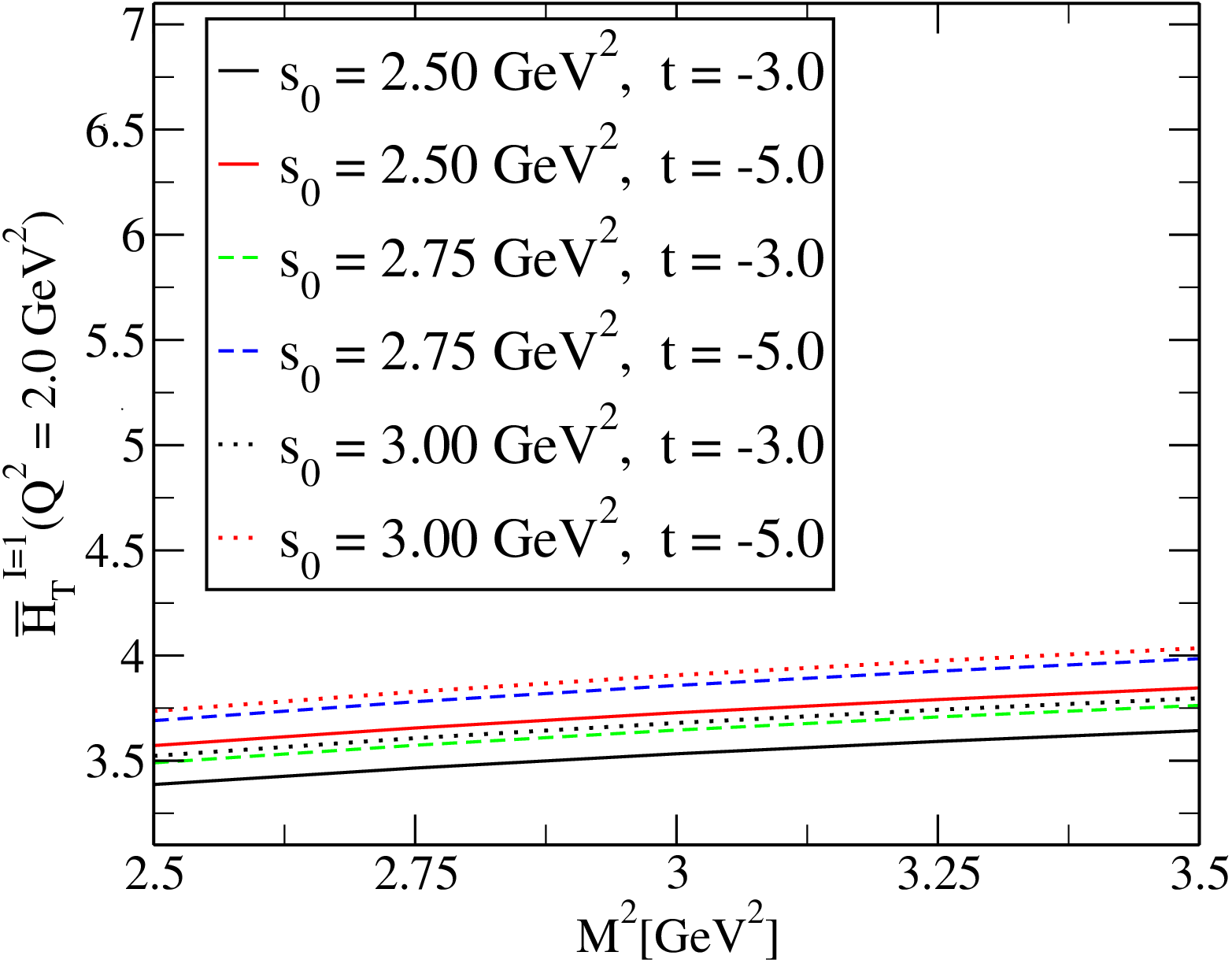}}~~~
 \subfloat[]{ \includegraphics[width=0.35\textwidth]{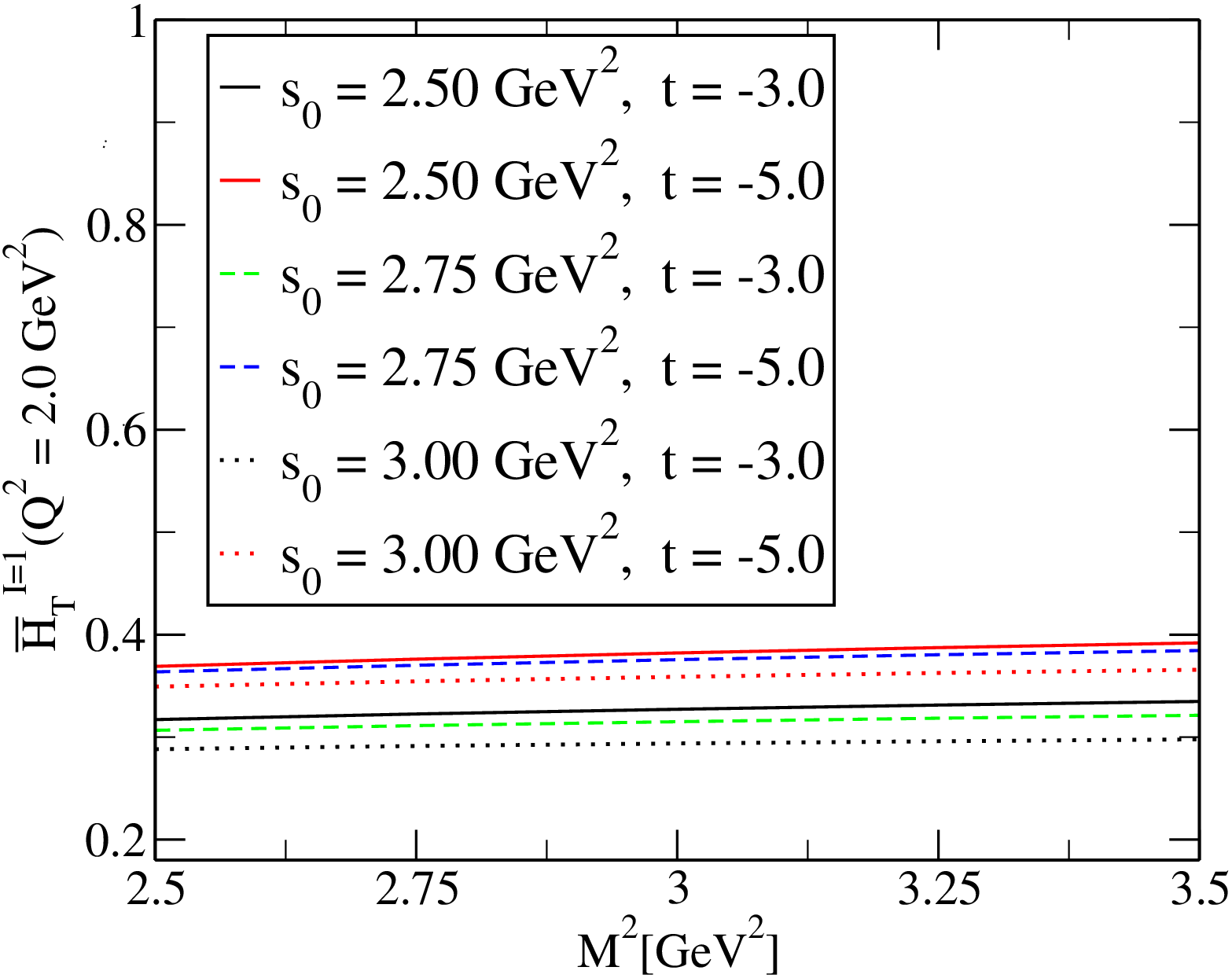}}~~~\\
 \caption{The dependence of the isovector tensor form factors of the $N(1535) \rightarrow N$ transition on $M^2$ at $Q^2$ = 2.0~GeV$^2$ and different values of $ s_0 $ and $ t $ at their working windows.  (a), (c) and (e) for LCSR-I,
and; (b), (d) and (f) for LCSR-II.}
 \label{isovectorMsqfigs}
  \end{figure}

 \begin{figure}[t]
\centering
  \subfloat[]{ \includegraphics[width=0.35\textwidth]{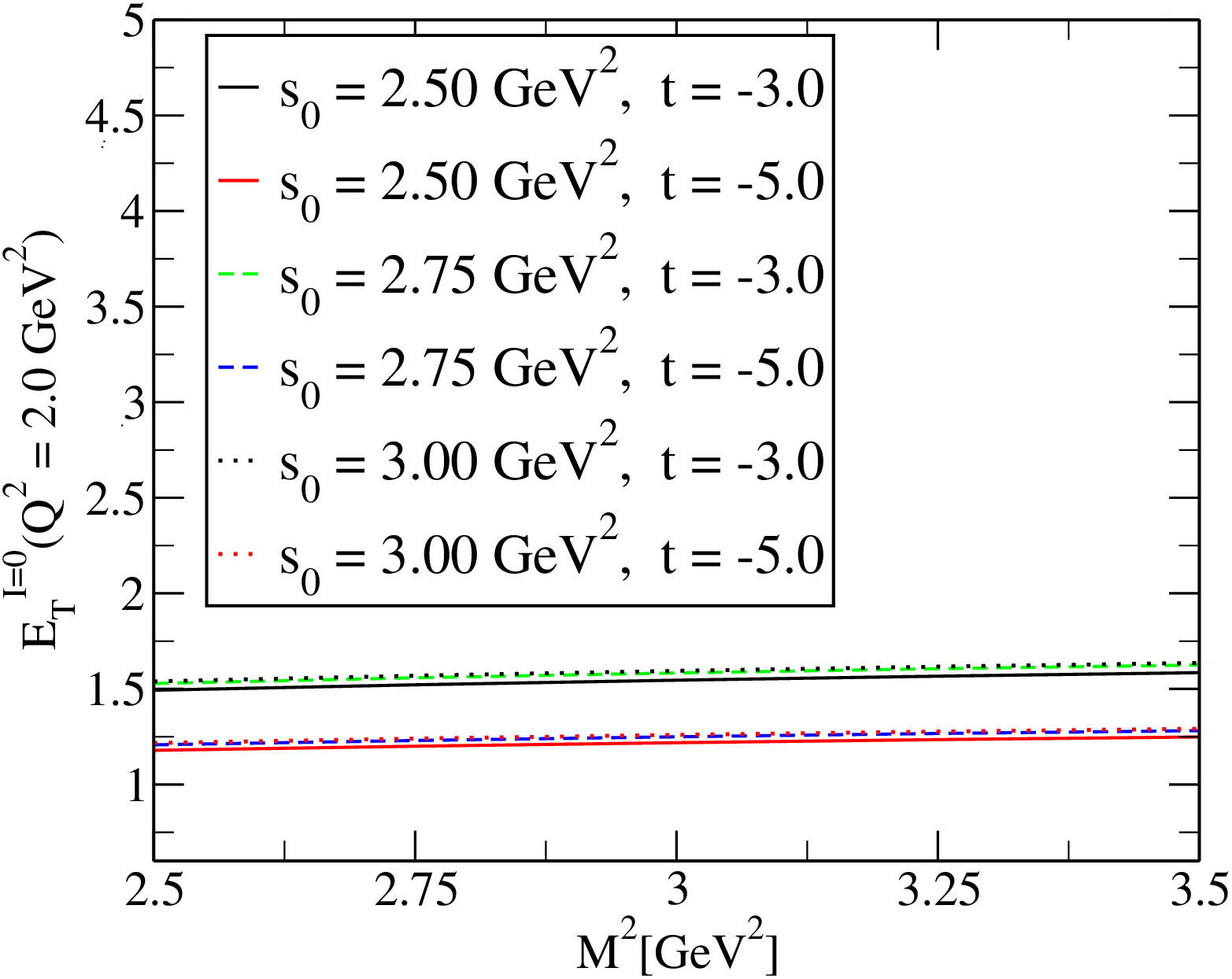}}~~~
 \subfloat[]{ \includegraphics[width=0.35\textwidth]{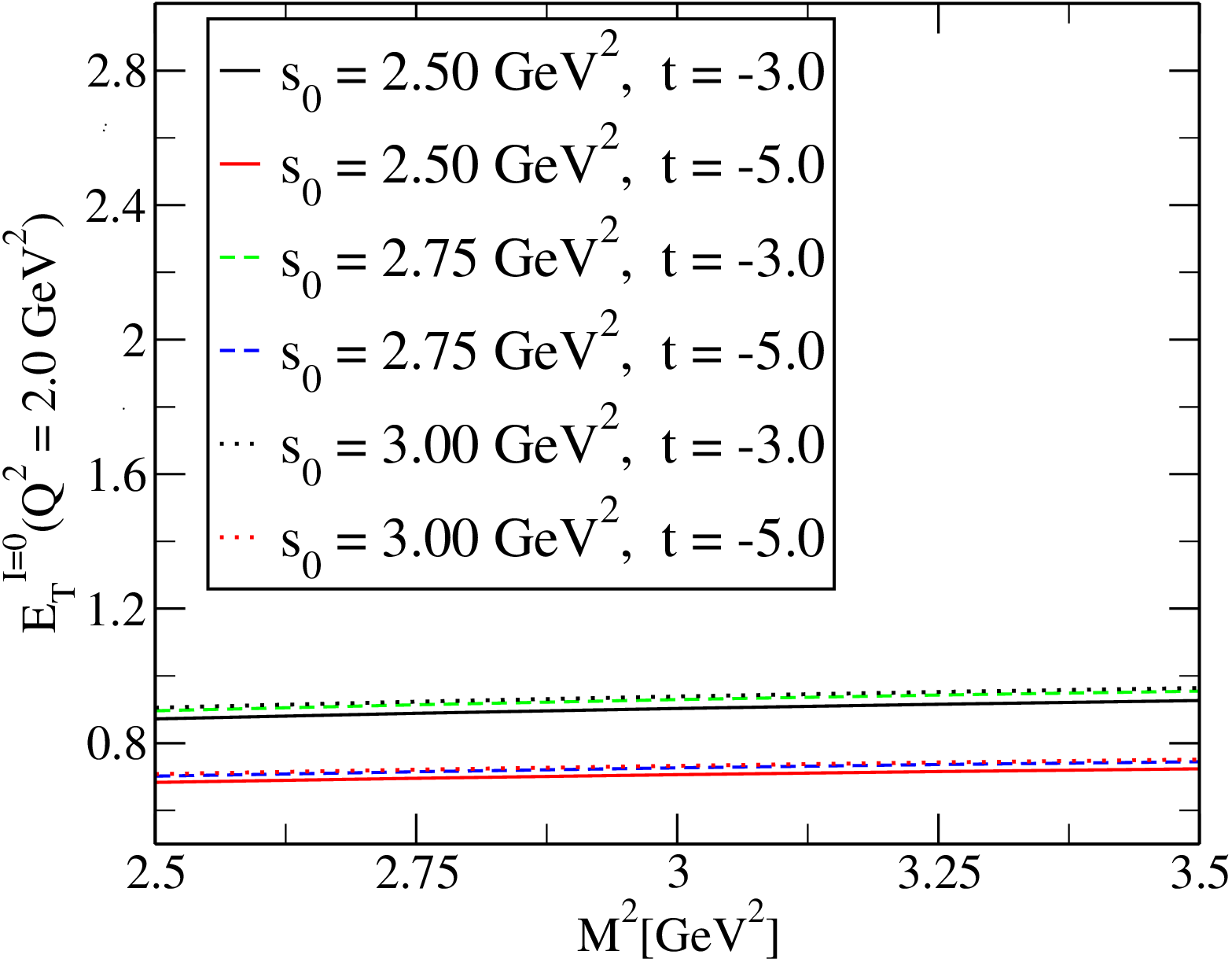}}~~~\\
  \subfloat[]{ \includegraphics[width=0.35\textwidth]{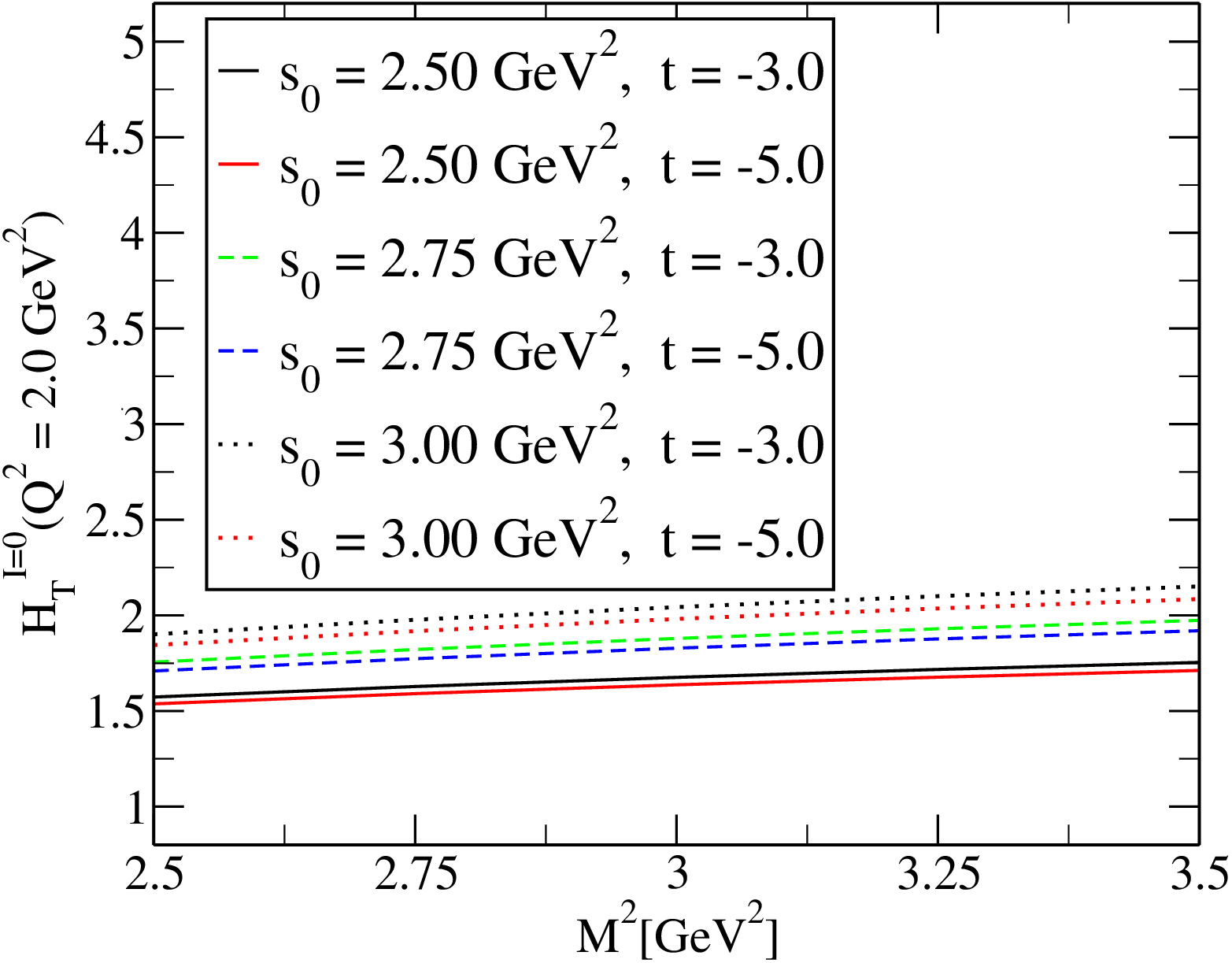}}~~~
 \subfloat[]{ \includegraphics[width=0.35\textwidth]{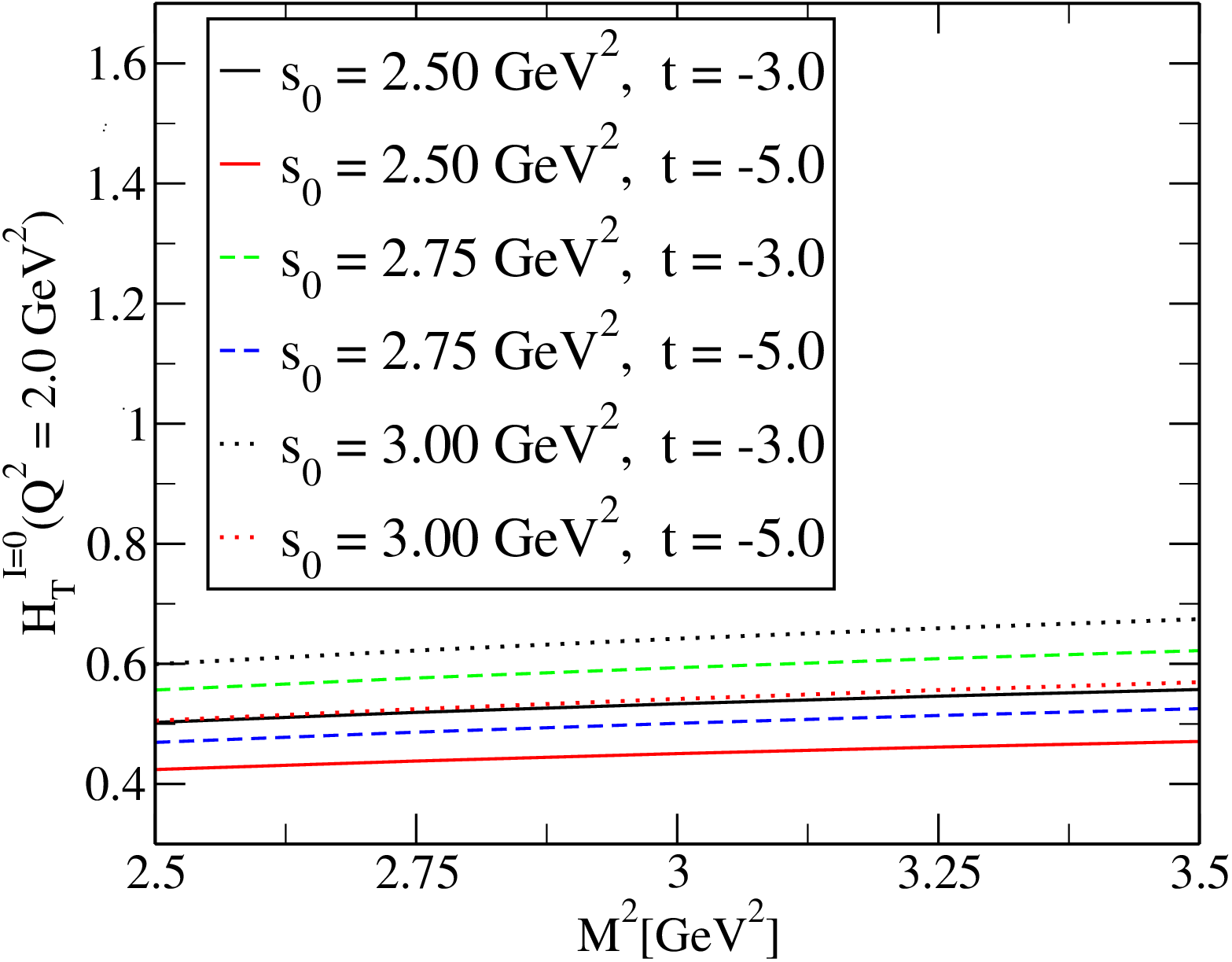}}~~~\\
 \caption{The dependence of the isoscalar tensor form factors of the $N(1535) \rightarrow N$ transition on $M^2$ at $Q^2$ = 2.0~GeV$^2$ and different values of $ s_0 $ and $ t $ at their working windows.  (a) and (c) for LCSR-I,
and; (b) and (d) for LCSR-II.}
 \label{isoscalarMsqfigs}
  \end{figure}


 \begin{figure}[t]
\centering
  \subfloat[]{ \includegraphics[width=0.35\textwidth]{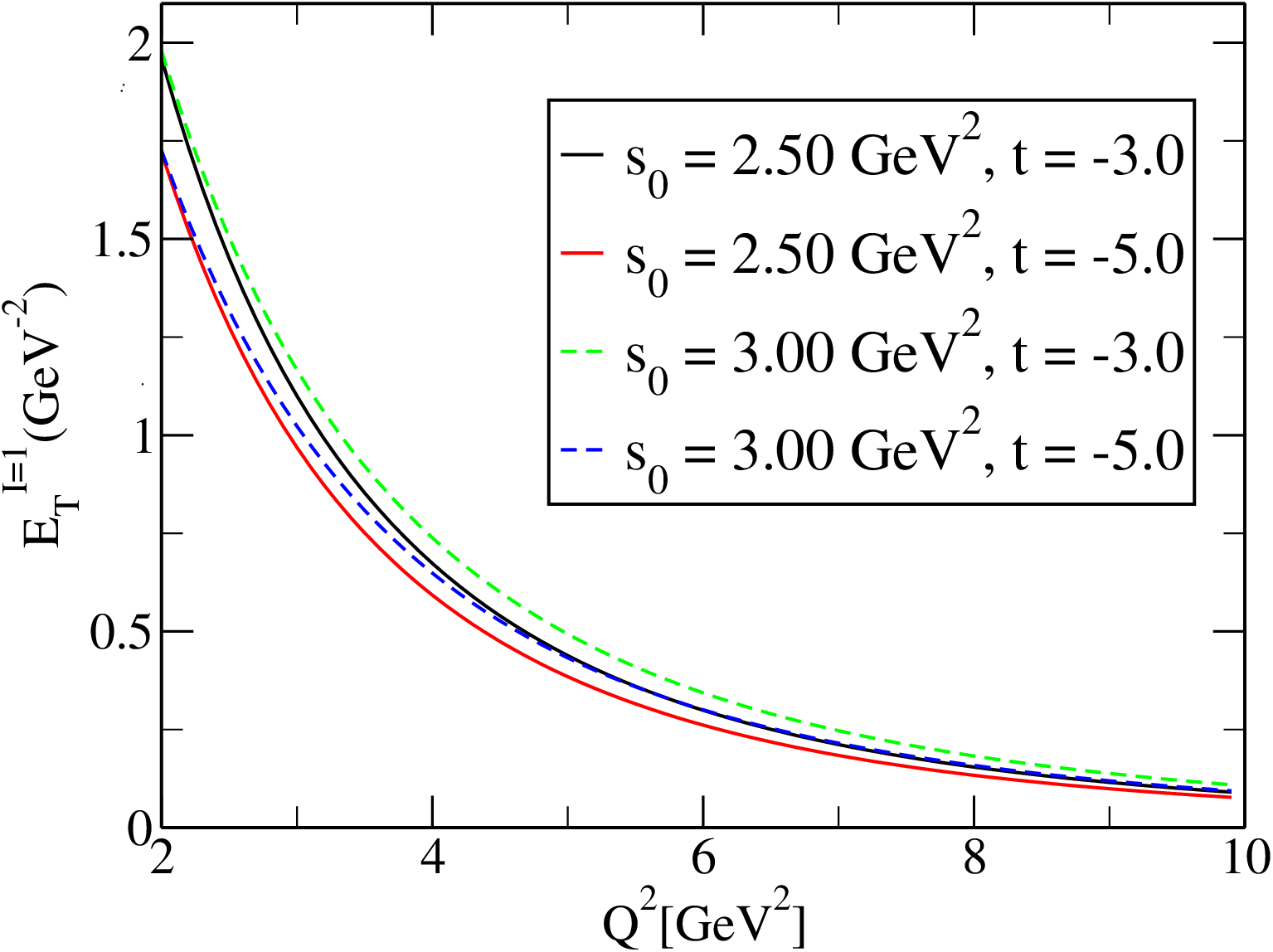}}~~~
 \subfloat[]{ \includegraphics[width=0.35\textwidth]{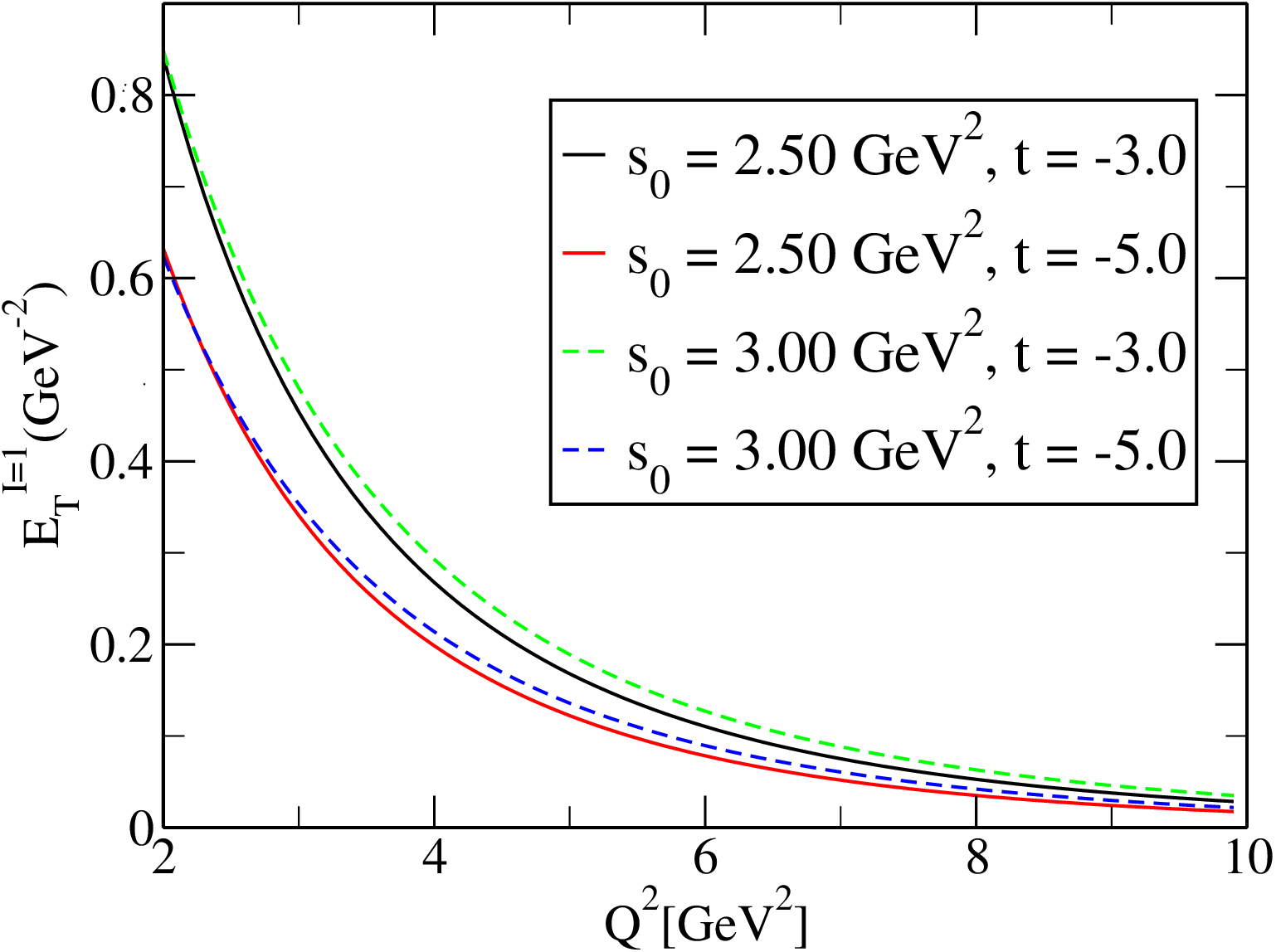}}~~~\\
  \subfloat[]{ \includegraphics[width=0.35\textwidth]{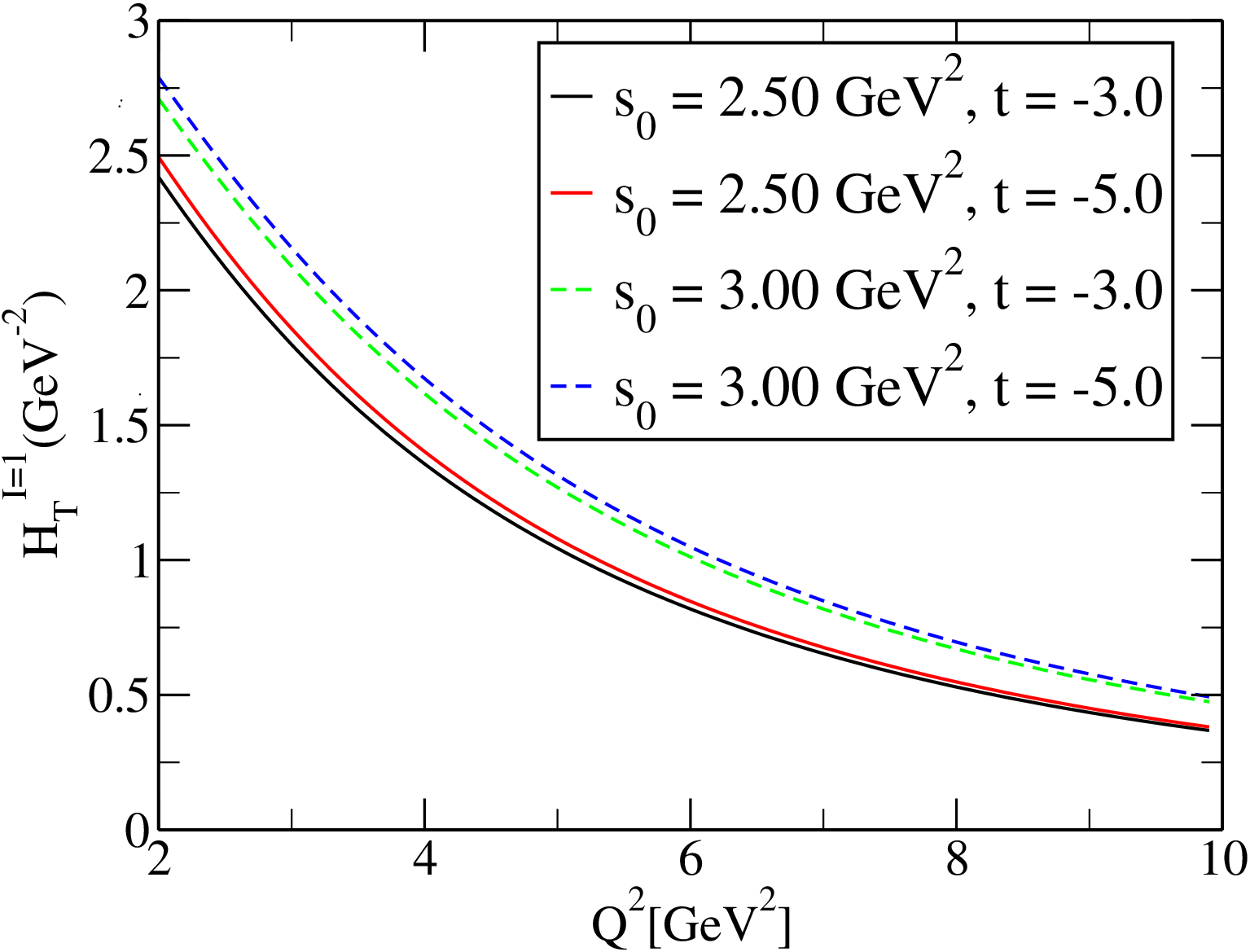}}~~~
 \subfloat[]{ \includegraphics[width=0.35\textwidth]{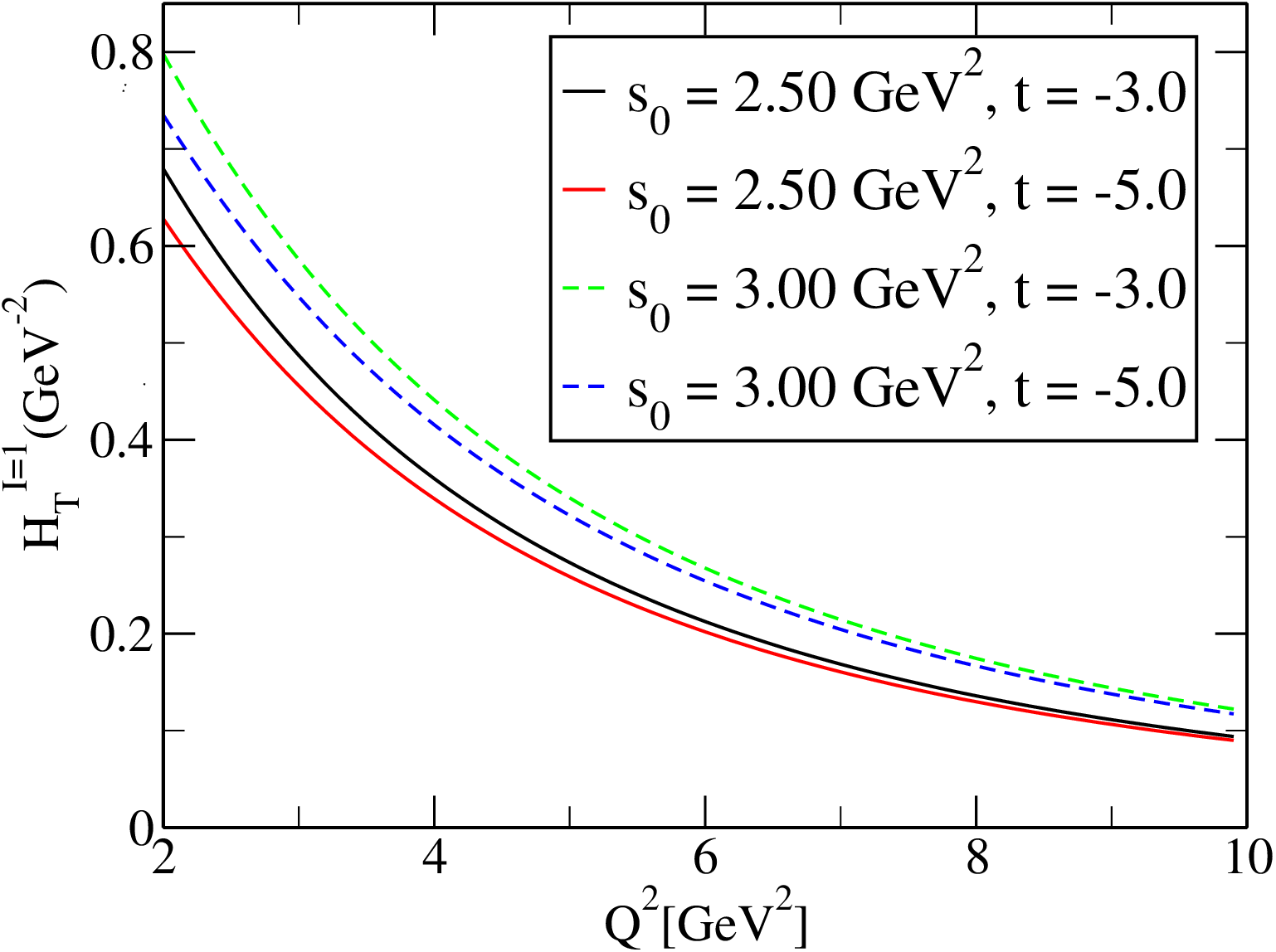}}~~~\\
  \vspace{0.2cm}
   \subfloat[]{ \includegraphics[width=0.35\textwidth]{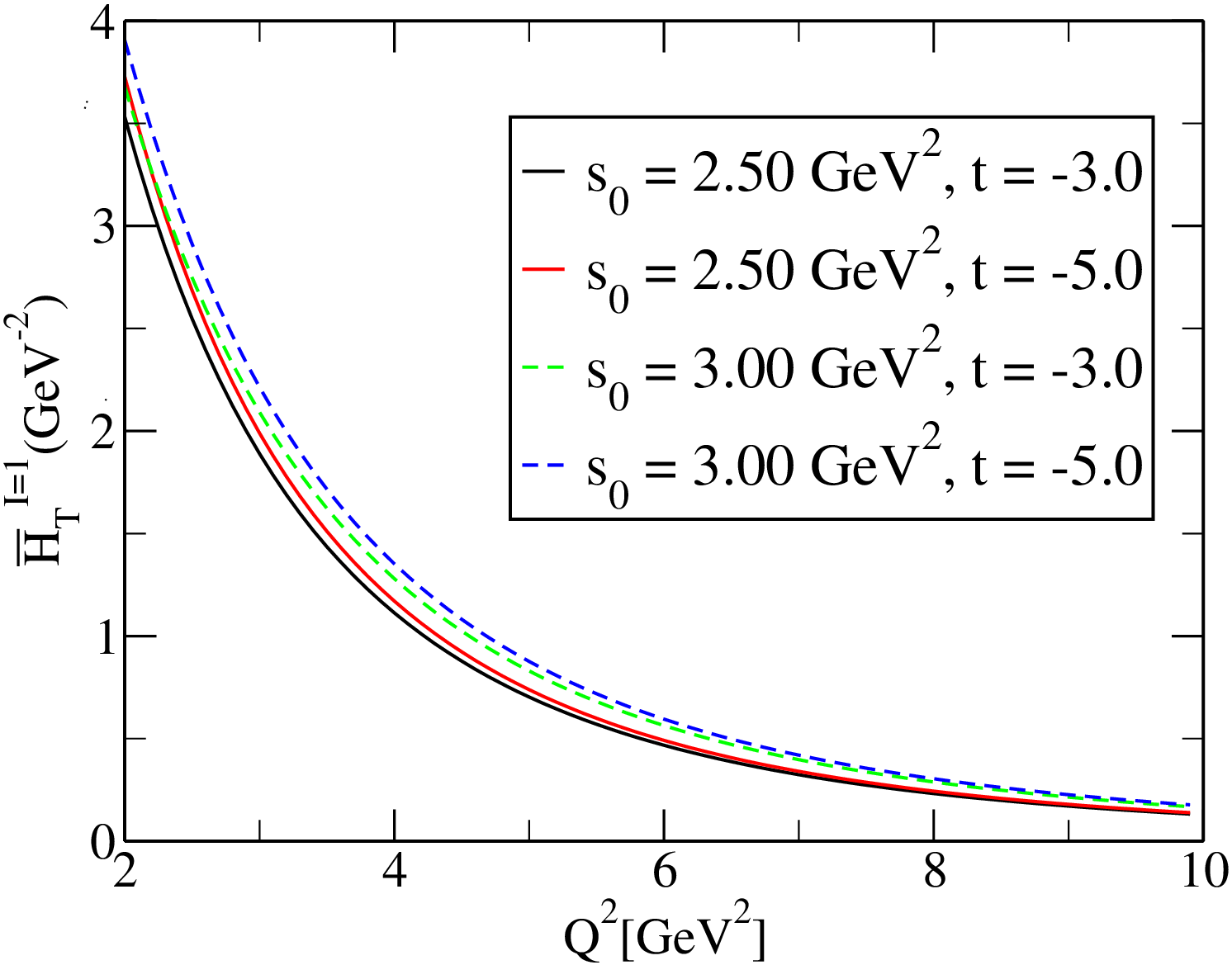}}~~~
 \subfloat[]{ \includegraphics[width=0.35\textwidth]{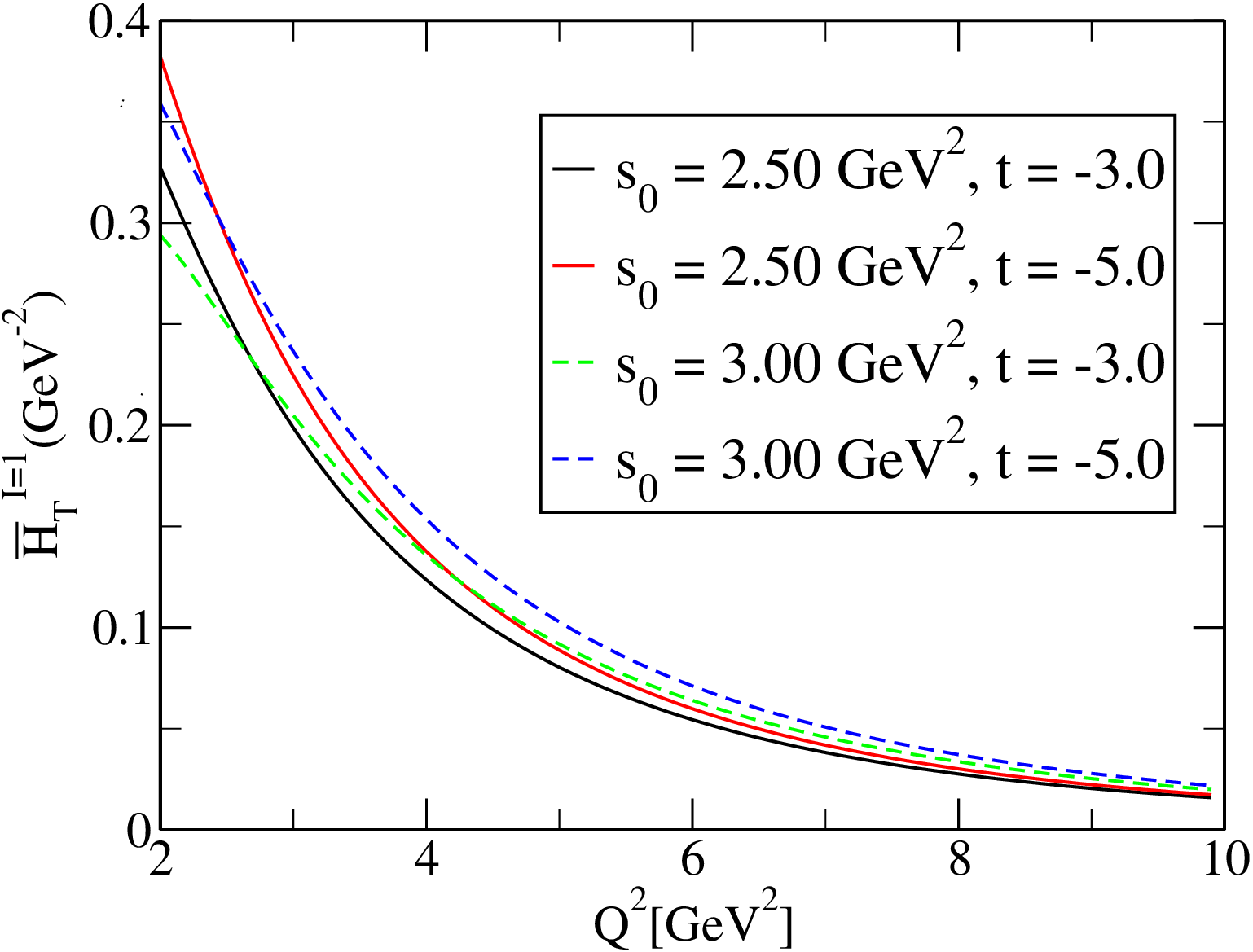}}~~~\\
 \caption{The dependence of the isovector tensor form factors of the $N(1535) \rightarrow N$ transition on $Q^2$ at $M^2$ = 3.00~GeV$^2$ and different values of $ s_0 $ and $ t $ at their working windows. (a), (c) and (e) for LCSR-I, and; (b), (d) and (f) for LCSR-II.}
 \label{isovectorQsqfigs}
  \end{figure}

 \begin{figure}[t]
\centering
  \subfloat[]{ \includegraphics[width=0.35\textwidth]{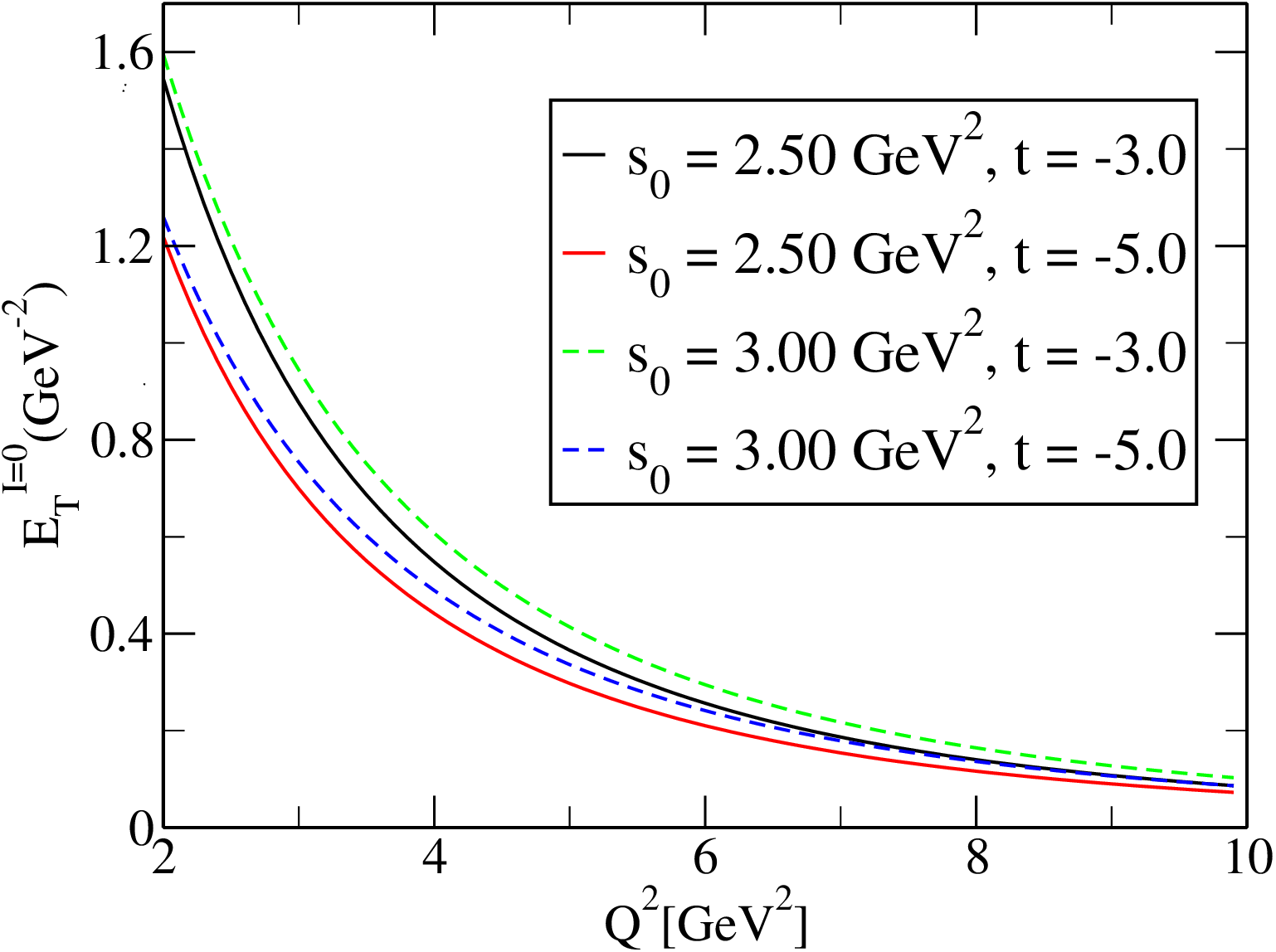}}~~~
 \subfloat[]{ \includegraphics[width=0.35\textwidth]{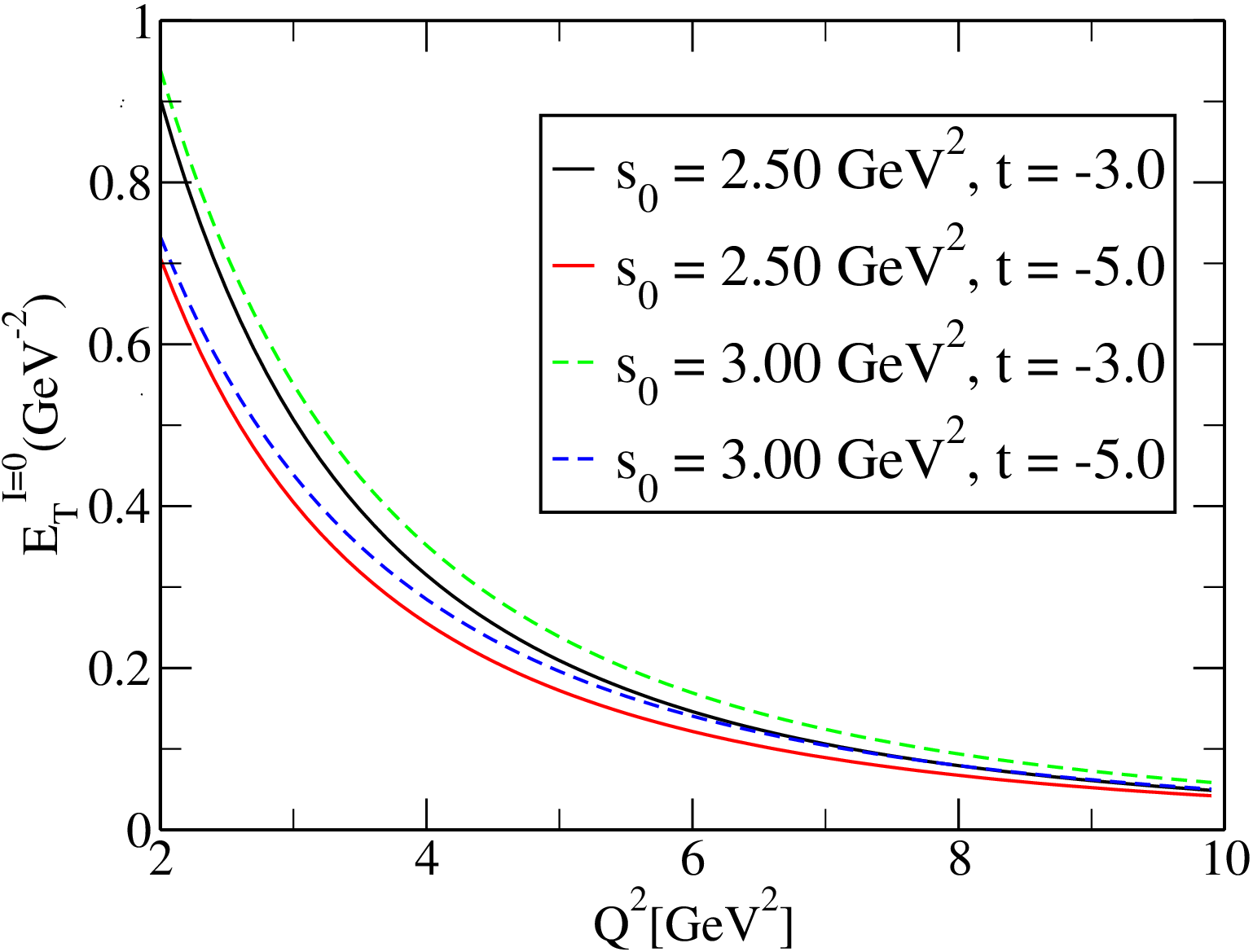}}~~~\\
  \subfloat[]{ \includegraphics[width=0.35\textwidth]{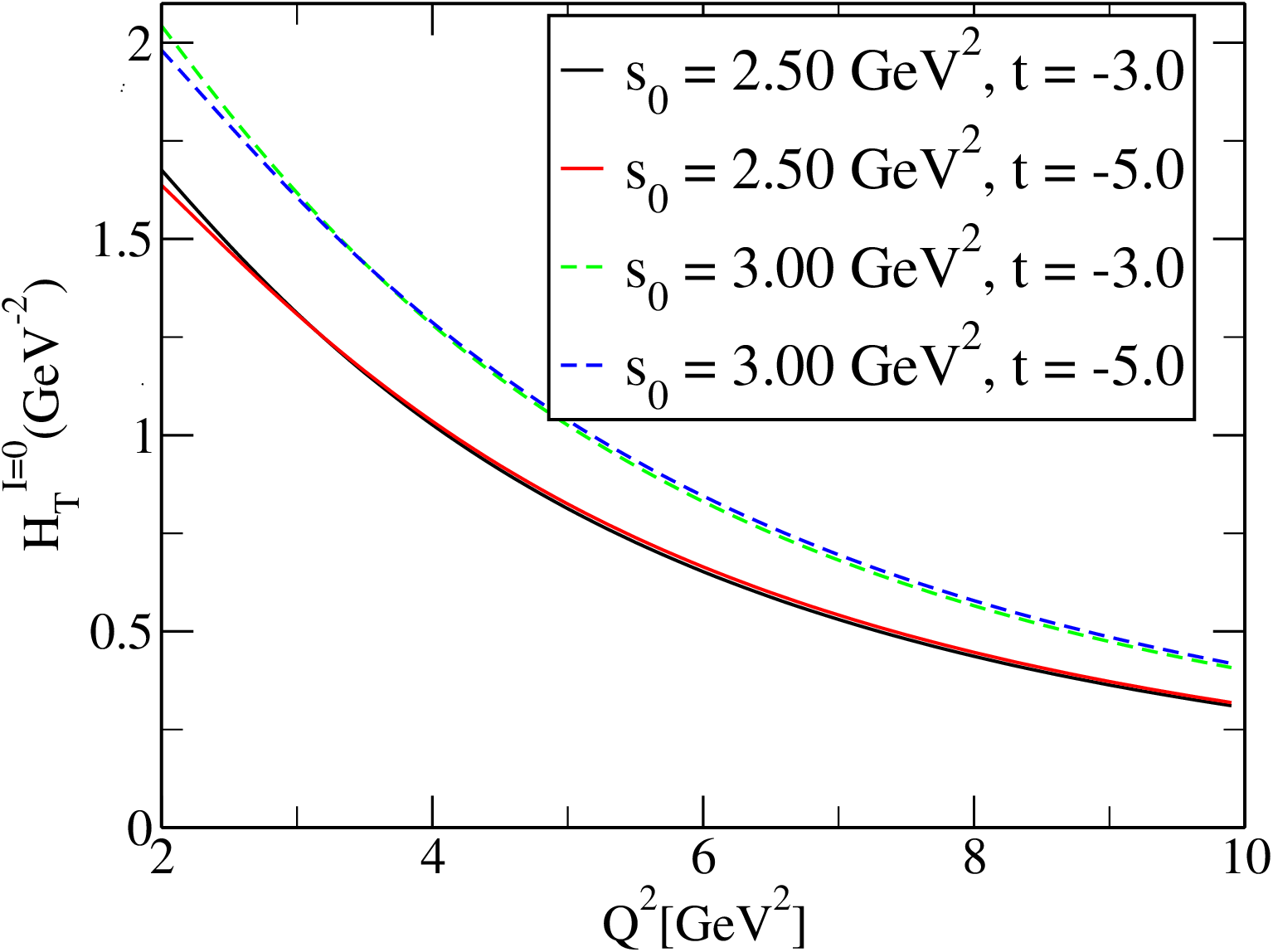}}~~~
 \subfloat[]{ \includegraphics[width=0.35\textwidth]{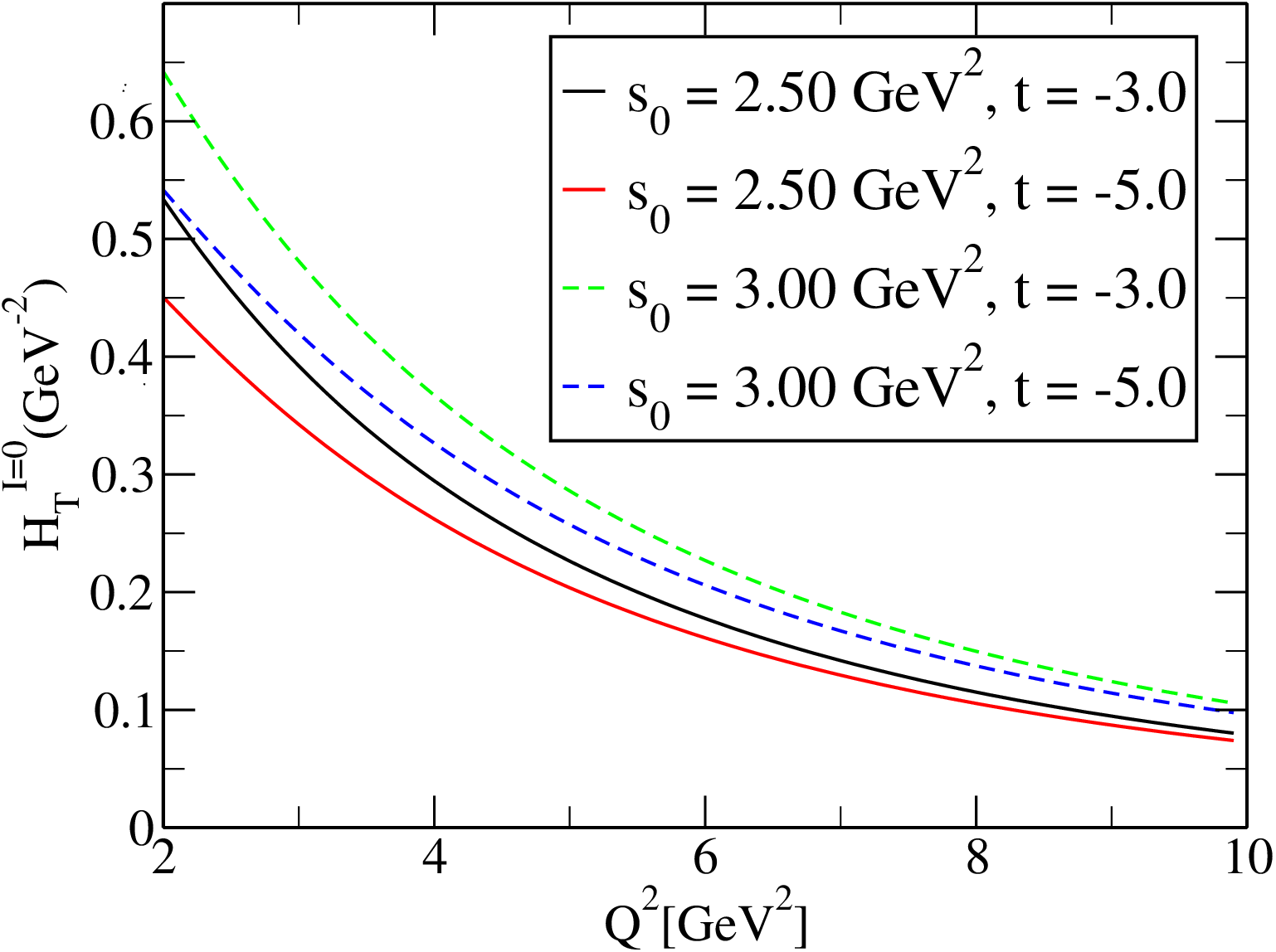}}~~~\\
 \caption{The dependence of the isoscalar tensor form factors of the $N(1535) \rightarrow N$ transition on $Q^2$ at $M^2$ = 3.00~GeV$^2$ and different values of $ s_0 $ and $ t $ at their working windows.  (a) and (c) for LCSR-I, and; (b) and (d) for LCSR-II.}
 \label{isoscalarQsqfigs}
  \end{figure}

  \end{widetext}


\end{document}